\def\draftversion{false}
\RequirePackage{ifthen}
\ifthenelse{\equal{\draftversion}{true}}{
  \documentclass[aps,prb,10pt,galley,amsmath,amssymb,showpacs,
                 superscriptaddress]{revtex4-1}
}{
  \documentclass[aps,prb,10pt,twocolumn,showpacs,
                 superscriptaddress]{revtex4-1}
}

\usepackage{times}
\usepackage{amsmath}
\usepackage{amssymb }
\usepackage{graphicx}
\usepackage[usenames]{color} 
\usepackage{bm} 
\usepackage{subfigure}

\def\Z2{$\mathbb{Z}_2$}
\def\bisb{(Bi$_{1-x}$Sb$_{x})_2$Se$_3$}
\def\biin{(Bi$_{1-x}$In$_{x})_2$Se$_3$}
\def\bise{Bi$_2$Se$_3$}
\def\inse{In$_2$Se$_3$}
\def\sbse{Sb$_2$Se$_3$}
\def\I{\uppercase\expandafter{\romannumeral 1}}
\def\II{\uppercase\expandafter{\romannumeral 2}}
\def\III{{\uppercase\expandafter{\romannumeral 3}}}
\def\IV{{\uppercase\expandafter{\romannumeral 4}}}
\def\Im{\operatorname{Im}}

\begin{document}

\title{Topological phase transitions in \biin\ and \bisb}
\date{today}

\author{Jianpeng Liu}
\affiliation{ Department of Physics and Astronomy, Rutgers University,
 Piscataway, NJ 08854-8019, USA }

\author{David Vanderbilt}
\affiliation{ Department of Physics and Astronomy, Rutgers University,
 Piscataway, NJ 08854-8019, USA }

\date{\today}

\begin{abstract}
We study the phase transition from a topological to a normal
insulator with concentration $x$ in (Bi$_{1-x}$In$_{x})_2$Se$_3$
and (Bi$_{1-x}$Sb$_{x})_2$Se$_3$ in the Bi$_2$Se$_3$ crystal
structure.  We carry out first-principles calculations
on small supercells, using this information to build
Wannierized effective Hamiltonians for a more realistic treatment
of disorder.
Despite the fact that the spin-orbit coupling (SOC) strength is
similar in In and Sb, we find that the critical concentration
$x_{\rm c}$ is much smaller in (Bi$_{1-x}$In$_{x})_2$Se$_3$ than
in (Bi$_{1-x}$Sb$_{x})_2$Se$_3$.  For example, the direct
supercell calculations suggest that $x_{\rm c}$ is
below $12.5\%$ and above 87.5$\%$ for the two alloys respectively.
More accurate results are obtained from
realistic disordered calculations, where the topological
properties of the disordered systems are understood from a
statistical point of view.  Based on these calculations,
$x_c$ is around $17\%$ for (Bi$_{1-x}$In$_{x})_2$Se$_3$, but
as high as 78\%-83\% for (Bi$_{1-x}$Sb$_{x})_2$Se$_3$. In
(Bi$_{1-x}$Sb$_{x})_2$Se$_3$, we find that the phase transition
is dominated by the decrease of SOC, with a crossover
or ``critical plateau'' observed from around
78$\%$ to 83$\%$. On the other hand, for
(Bi$_{1-x}$In$_{x})_2$Se$_3$, the In 5$s$ orbitals suppress the
topological band inversion at low impurity concentration, therefore
accelerating the phase transition. In (Bi$_{1-x}$In$_{x})_2$Se$_3$
we also find a tendency of In atoms to segregate.
\end{abstract}


\maketitle


\def\scr{\scriptsize}
\ifthenelse{\equal{\draftversion}{true}}{
  \marginparwidth 2.7in
  \marginparsep 0.5in
  \newcounter{comm} 
  \def\commnext{\stepcounter{comm}}
  \def\commtext{{\bf\color{blue}[\arabic{comm}]}}
  \def\commmar{{\bf\color{blue}[\arabic{comm}]}}
  \def\dvm#1{\commnext\marginpar{\small DV\commmar: #1}\commtext}
  \def\jlm#1{\commnext\marginpar{\small JPL\commmar: #1}\commtext}
  \def\mlab#1{\marginpar{\small\bf #1}}
  \def\tnewpage{\newpage\marginpar{\small Temporary newpage}}
}{
  \def\dvm#1{}
  \def\jlm#1{}
  \def\mlab#1{}
  \def\tnewpage{}
}

\section{Introduction}

Topological aspects of quantum systems has been an exciting area
in condensed-matter physics since the discovery of the integer
quantum Hall effect (IQHE)\cite{IQH-prl80,TKNN} and the first
proposal of a 2D Chern insulator.\cite{Haldane-model} Both the
IQHE and the 2D Chern insulators are characterized by a quantized
Hall conductance and the presence of gapless edge modes that
are topologically protected by a non-zero Chern number. In 2005,
a topological classification was also found to apply to
spinful systems with SOC and
time-reversal symmetry, defining a topologically non-trivial 2D state
known as a quantum spin Hall (QSH)
insulator.\cite{kane-prl05-a,kane-prl05-b} A QSH insulator also
possesses gapless edge states that always
cross at one of the time-reversal invariant momenta (TRIM) in
the 1D edge Brillouin zone (BZ). In a 2D Chern insulator,
the chiral gapless edge modes can be interpreted in terms of the charge
accumulation at one end of a truncated 1D system during an
adiabatic periodic evolution. The spin-polarized edge modes
in a QSH insulator can be interpreted in a similar way, except
that charges with opposite spin characters are pumped in opposite
directions and accumulated on opposite ends.\cite{fu-prb06} This pumping
process can be classified by a
new topological index, known as the \Z2\ index, which guarantees
the robustness of the edge modes of a QSH insulator to weak
time-reversal invariant perturbations.

The definition of the \Z2\ index was later generalized from 2D
to 3D crystals.\cite{fu-prl07,moore-prb07} In 3D systems, there
is one strong \Z2\ index, which is odd when the number of Dirac
cones on the surface is odd; when it is even, the other three
indices characterize the weak TIs, specifying how these gapless
surface states are distributed among the TRIM in the 2D surface BZ.

A non-trivial bulk topological index is usually connected with a
non-trivial ``topological gap'' resulting from band inversion. For
systems with inversion symmetry, the topological index can be
uniquely determined from the parities of the occupied states at
the TRIM in the BZ.\cite{fu-prb07}
Thus,  to drive an inversion-symmetric system from a normal insulator
(NI) to a TI, a strong SOC is usually needed to flip the valence-band
maximum (VBM) and conduction-band minimum (CBM) with opposite
parities at one of the TRIM. The band
gap after the topological band inversion is conventionally assigned
with a minus sign, to be distinguished from the ordinary band gap
in the \Z2-even case.
The scenario sketched above is exactly the mechanism in the
Bi$_2$Se$_3$ class
of TIs.\cite{zhang-np09,cxliu-prb10,hsieh-prl09,xia-np09,
hsieh-science09,wray-np10,hsieh-nature09,chen-science09}
In Bi$_2$Se$_3$ with SOC turned off, the VBM and CBM states at $\Gamma$
are built from Se $4p$ and Bi $6p$ orbitals in such a way as to
have opposite parities.
When SOC is turned on, the previous VBM is pushed up into the conduction
bands, leading to an exchange of parities and a non-trivial
\Z2\ index. As long as the inverted band gap remains
and time-reversal symmetry is preserved, a single Dirac cone exhibiting
a helical spin texture is guaranteed to exist at $\overline{\Gamma}$
in the surface BZ.  For some useful recent reviews, see
Refs.~\onlinecite{kane-rmp10,zhang-rmp11,yao-review12}.

Up to now, however, only a few pioneering
works\cite{ando-np11,xu-science11,chadov-cpa12} have focused on
the topological phase transition from the TI to the NI state
driven by non-magnetic substitution, and while the general picture of
such a transition seems obvious, details remain unclear.
In the simplest picture, one would expect the band gap of a TI
to decrease linearly as a lighter element with weaker SOC is
substituted, and the phase transition would occur when the bulk
gap is closed. However, on a closer look, many questions arise.
For example, the bandstructures of known TIs are mostly dominated
by $p$ orbitals, but what happens if the substituted element
includes different valence orbitals such as $s$ or $d$ orbitals?
More fundamentally, translational symmetry is lost for a
randomly substituted system.  In this case, how should one
determine the topological properties of a system in which wavevector
$\mathbf k$ is no longer a good quantum number, and what signature
indicates the presence of a TI state?
These questions focus on two aspects that are not
taken into account in the simplest linear band-closure picture:
the effects of impurities with different orbital character,
and the effects of disorder.

These issues arise, in particular, for the substitution of In
into \bise, one of the best-known TI systems.
Recently, several experimental groups have reported a surprisingly low
critical concentration $x_{c}$ of about 5\%
in \biin, much lower than would be expected from a linear
band-closure picture, thus challenging the usual understanding of the
phase-transition behavior of TIs by non-magnetic
doping.\cite{oh-prl12,armitage1} These experiments
motivated our theoretical studies of the \biin\ system. Moreover,
to separate the effects of In $5s$
orbitals from a simple weakening of the effective SOC, we
also study \bisb. Here Sb has the same orbital character as Bi,
lying directly above it in the Periodic Table,
but shares the weaker intrinsic SOC strength of In because
their atomic numbers are very close in magnitude.

We first study the solid-solution systems by constructing small
supercells with different impurity configurations. For each supercell
configuration, the strong \Z2\ index and surface states
are computed using Wannier-interpolation techniques,\cite{MLWF-rmp}
which also allow us to test the effect of artificially removing
the In $5s$ orbitals from the calculation.
Next, we study the the effects of disorder more realistically by
constructing a large supercell of pure \bise\ acting as reference system,
making random substitutions of In or Sb on the Bi sites,
and calculating the disorder-averaged
spectral functions.\cite{weiku1,weiku2} We further propose an
approach in which we compute ``\Z2\text{-}index statistics''
in order to determine the
topological properties of disordered systems from a statistical
point of view.

Based on our results, the \bisb\ system
is well described by the linear band-closure picture
with a high critical concentration $x_c$, because
the orbital character of the host and dopant are the same and
the disorder effect is thus rather weak. We also observe a
``critical plateau'' in the Sb-substituted system, where the critical Dirac
semimetal phase remains robust from about $x\approx$78\% to $x\approx$83\%, 
although it is difficult to test whether this may be a finite-size effect due
to the limited numerical accuracy in our calculations.
In In-substituted Bi$_2$Se$_3$, on the other hand, the disorder effects are
quite strong, and the presence of In $5s$ orbitals rapidly drives
the system into the NI state even at very low impurity concentrations.
A tendency of segregation of In atoms has been observed for
\biin, and may play an important role.

The paper is organized as follows. In Sec.~\II\ the lattice
structures, notations and the details of first-principles
calculations are introduced. In Sec.~\III\ we present the main
results of this paper. First we summarize the results from
the direct first-principles superlattice calculations, and determine the
critical points and the influence of In $5s$ orbitals by computing
the bulk \Z2\ index and by calculating surface states. Then, the critical
points of the two solid-solution systems are further determined by looking
at the disordered spectral functions, and the topological behaviors
are understood from a statistical point of view. Finally we summarize
in Sec.~\IV.

\section{Preliminaries}

\subsection{Structures of bulk material and superlattices}
\label{sec:prelim_lattice}

\begin{figure}
\centering
\includegraphics[width=8.7cm]{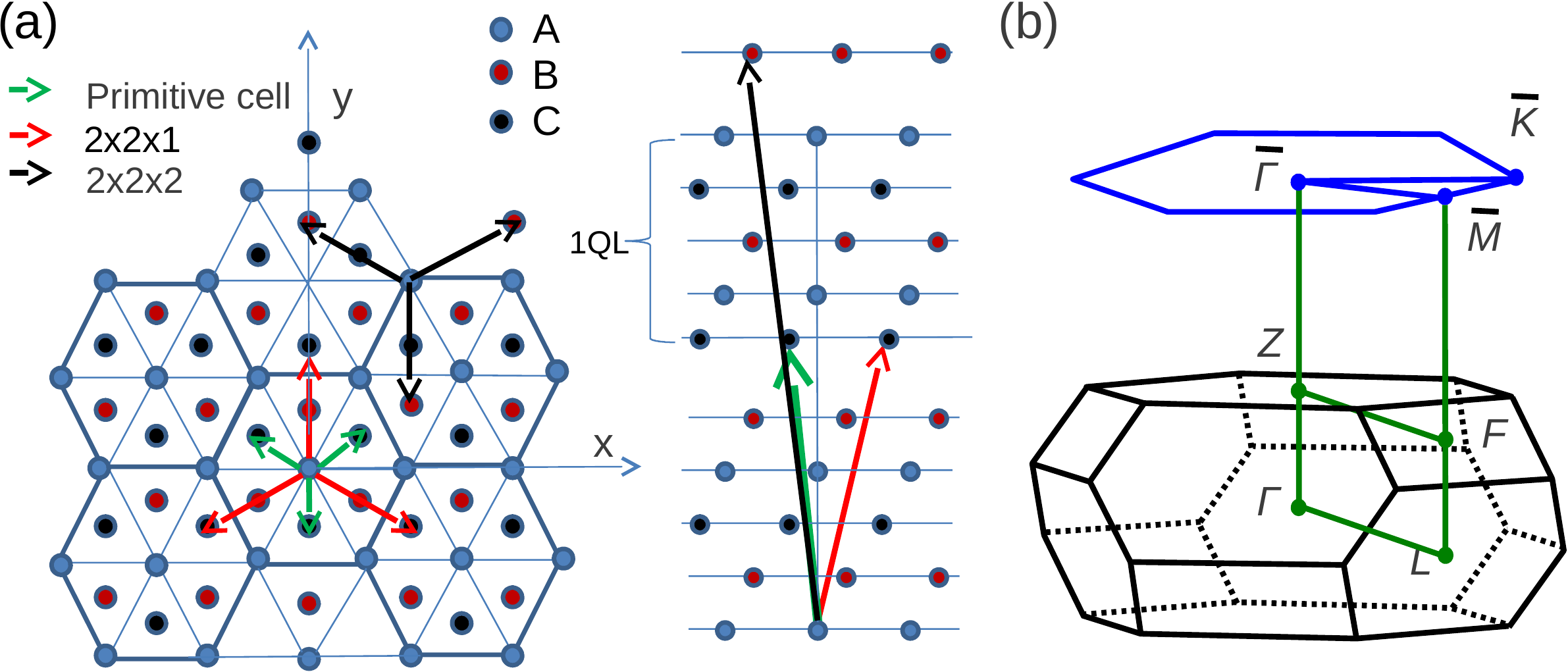}
\caption{(a) Lattice vectors of \bise\ primitive cell,
2$\times$2$\times$1 supercell, and 2$\times$2$\times$2
supercell.  (b) Corresponding bulk Brillouin zone and its
surface projection.}
\label{fig:latt}
\end{figure}

As shown in Fig.~\ref{fig:latt},
the crystal structure of Bi$_2$Se$_3$ is
rhombohedral. The crystal has a layered structure along the $z$ direction
with five atoms per primitive cell. The five 2D monolayers made by
repeating the primitive cell in the $x$ and $y$ directions form a
quintuple layer (QL).
In each QL, there are two equivalent Se atoms located
at the top and bottom of the QL, two equivalent Bi atoms inside
those, and another
central Se atom. Seen from the top, each monolayer forms a 2D
triangular lattice, and these triangular planes are stacked along
the $z$ direction in the order $A-B-C-A-B-C$..., where $A$, $B$ and $C$
represent the three different high-symmetry sites. Both Bi$_2$Se$_3$ and
$\beta$-phase In$_2$Se$_3$ have a rhombohedral structure belonging
to the $R\overline{3}$m space group,
but their lattice parameters are
slightly different. The in-plane hexagonal lattice parameter is
$a\!=\!4.138$\,\AA\ for \bise\ but $a=4.05$\,\AA\ for \inse, and the
height of a QL is $c\!=\!9.547$\,\AA\ for \bise\ compared with
$c\!=\!9.803$\,\AA\
for \inse.  The rhombohedral structure of \sbse\ does not exist
in nature, so for this case we relaxed both the lattice
parameters and atomic positions. After a complete relaxation,
we obtained $a\!=\!4.11$\,\AA\ and $c\!=\!10.43$\,\AA\ for \sbse.

To study the substitution problem from first-principles
calculations,
a 2$\times$2$\times$1 supercell based on the original Bi$_2$Se$_3$
crystal structure is built. The lattice vectors of the supercell
are shown in Fig.~\ref{fig:latt}. There are 20 atomic sites in such
a supercell, where eight of them are Bi sites.
Among all the possible configurations, we choose to investigate
the supercells with one, two, four, six and seven Bi atoms substituted
by impurities.
The (unique) configuration with $x\!=\!0.125$ is denoted as C$_{0.125}$.
For two or six impurities ($x\!=\!0.25$ and $x\!=\!0.75$), there are two
inequivalent configurations, with the two impurity (or remaining host)
atoms residing in different monolayers or in the same monolayer, which
we label as C$_{0.25}$ (C$_{0.75}$) and C$'_{0.25}$
(C$'_{0.75}$) respectively.
For four impurities, $x\!=\!0.5$, all impurities can be clustered in
one monolayer, labeled as C$''_{0.5}$, or three in one monolayer and
one in the other, denoted as C$'_{0.5}$, or the impurities can be equally
divided between top and bottom monolayers with inversion symmetry,
denoted as C$_{0.5}$.  Note that primes indicate more strongly
clustered configurations.

\subsection{First-principles methodology}
\label{sec:first-principles}

The first-principles calculations are carried out with the 
\textsc{Quantum ESPRESSO} package,\cite{QE-2009} with the PBE 
generalized gradient approximation (GGA) exchange-correlation 
func\-tion\-al\cite{pbe-1,pbe-2} and 
well-tested fully relativistic ultrasoft\cite{vanderbilt-prb90} 
and norm-conserving pseudopotentials. The ultrasoft pseudopotentials 
are from \textsc{Quantum ESPRESSO},\cite{qe-pseudopotentials} 
and the norm-conserving pseudopotentials are
from the OPIUM package.\cite{opium-web,opium-paper}
The ionic relaxations, ground-state energies, and densities of states
presented in Sec.~\ref{sec:superlattice}
are calculated with ultrasoft pseudopotentials, but we switched
to norm-conserving pseudopotentials for those
topics that required transformation to a Wannier representation
(see below).
The energy cutoff with ultrasoft pseudopotentials is 60\,Ry for In-substituted
Bi$_2$Se$_3$ supercells and 35\,Ry for Sb-substituted supercells.
The cutoff becomes larger for norm-conserving pseudopotentials,
specifically 65\,Ry
for In substitution and 55\,Ry for Sb substitution.
The BZ is sampled on
a 6$\times$6$\times$6 Monkhorst-Pack\cite{monkhorst-pack} $\mathbf k$ mesh
for the 2$\times$2$\times$1 supercells,
and 8$\times$8$\times$8 for the
primitive cell bulk materials. In our calculations, the lattice
parameters of the Sb- and In-substituted supercells are fixed, 
taken as a linear interpolation of the Bi$_2$Se$_3$ and In$_2$Se$_3$ 
experimental lattice parameters according to the impurity concentration $x$,
and the internal coordinates of the atoms are fully relaxed.
We do not relax the lattice vectors because the coupling between two QLs
is at least partially of van der Waals type, so that the standard GGA
does not give a good estimate
of the lattice constants, especially the one in the $z$ direction.

To investigate the topological properties of these supercells,
we calculate both the bulk \Z2\ indices and the surface states
using the Wannier-interpolation technique. More
specifically,  we use the Wannier90 package to generate Wannier functions
(WFs) from the outputs of standard first-principles
calculations.\cite{wannier90} Wannier90 can optionally generate
maximally localized WFs,\cite{MLWF-1,MLWF-2} and in any case
reports the Wannier charge centers, their spreads, and
the real-space Hamiltonian matrix elements of an effective
tight-binding (TB) model in the WF basis.
This information is often very
useful in studying the bonding mechanism of materials, as well as
for calculating topological indices,
computing surface and interface states, treating disorder, etc.

It should be noted that the TB models constructed
from Wannier90 are realistic in the sense that the Wannier-inter\-po\-lated
bandstructures reproduce the first-principle bandstructures
essentially exactly within a certain energy window. This ``frozen
window'' is chosen to extend from 3\,eV below the Fermi level to
3\,eV above the Fermi level in our calculations. In addition to the
frozen window,
there is also an outer energy window outside which the Bloch eigenstates
will not be included in generating the WFs.  The outer
window varies in our calculations depending on the system, but
typically covers a total range of 17-22\,eV and includes
all the valence $p$ bands as well as In valence $s$ bands when
present. For example, for \bise we construct 30 spinor WFs per
primitive cell, and two additional WFs
constructed from In valence $s$ orbitals would be added for each
substituted In atom.

\section{Results and discussions}
\label{sec:superlattice}

\subsection{Ground-state energies and band gaps}
\label{sec:energies}

\begin{figure}
\centering
\includegraphics[height=5.5cm,width=8cm]{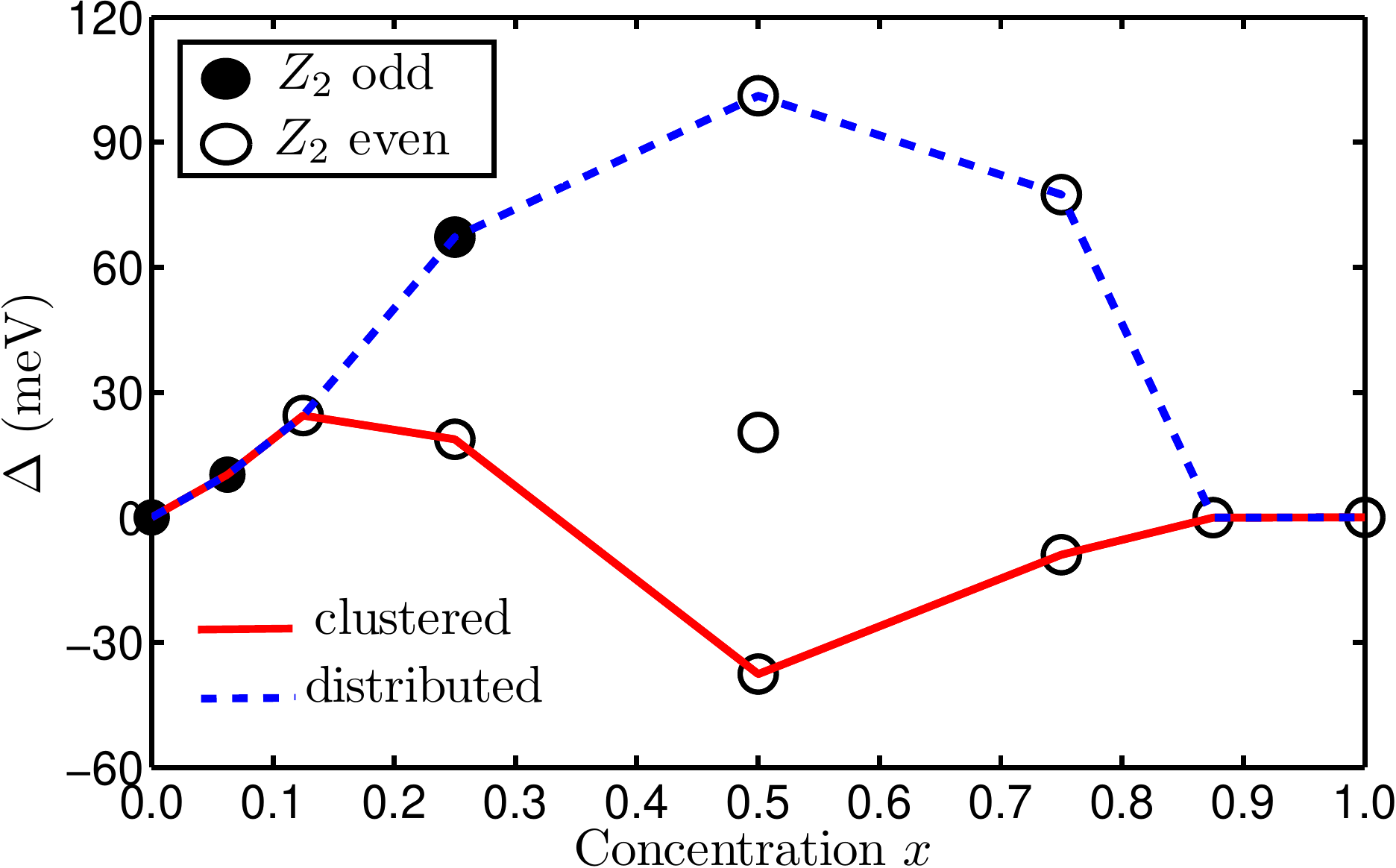}
\caption{Ground-state energies vs.~impurity
concentration $x$ for (Bi$_{1-x}$In$_{x})_2$Se$_3$ supercells.
Here $\Delta=E_{g}(x)-(1-x)E_1+x E_{2}$, where $E_{g}(x)$,
$E_1$ and $E_2$ are the ground-state energies per 5-atom cell
for the alloy supercell, host material, and dopant material,
respectively.  Filled and open circles denote \Z2-odd and
even states respectively.  Solid (red) and dashed (dark blue)
lines follow the most and least In-clustered configurations
respectively.}
\label{fig:In-supercell}
\end{figure}

We begin by discussing our results for In-substituted supercells
representing \biin.
The ground-state energies for supercells with different In impurity
configurations are shown in Fig.~\ref{fig:In-supercell}.
Open and closed circles represent topologically normal
and \Z2-odd cases respectively (see Sec.~\ref{sec:Z2}).  For
concentrations $0.25\le x \le 0.75$ there are two or more
inequivalent configurations of the 2$\times$2$\times$1 supercell
having the same concentration $x$.  Among these, the configurations
with lowest total energy are traced by the solid red line, and are found
to consist of ``clustered'' configurations in which the In impurities
tend to be first neighbors.  Conversely, those with the highest total
energies, indicated by the dark blue dashed
line, are those with the In atoms distributed most evenly
throughout the supercell.
For example, at $x\!=\!0.5$, the ground-state energy of the clustered
configuration (C$''_{0.5}$) is lower than that of the distributed
one (C$_{0.5}$) by 140\,meV per primitive unit cell,
and at $x\!=\!0.25$ the energy of C$'_{0.25}$ is lower than that of
C$_{0.25}$ by 50\,meV per primitive unit cell.  Thus we clearly
find a strong tendency of the In atoms to segregate and
cluster together. We also find that the \Z2\ index changes sign
at a critical concentration $x_c$ lying somewhere between
6.25\% and 12.5\%. (One may notice from Fig.~\ref{fig:In-supercell}
that the distributed configuration C$_{0.25}$ at $x\!=\!0.25$
is \Z2-odd, but since its energy
is so much higher, the significance of this is questionable.)

\begin{figure}
\centering
\includegraphics[width=8cm, height=6cm]{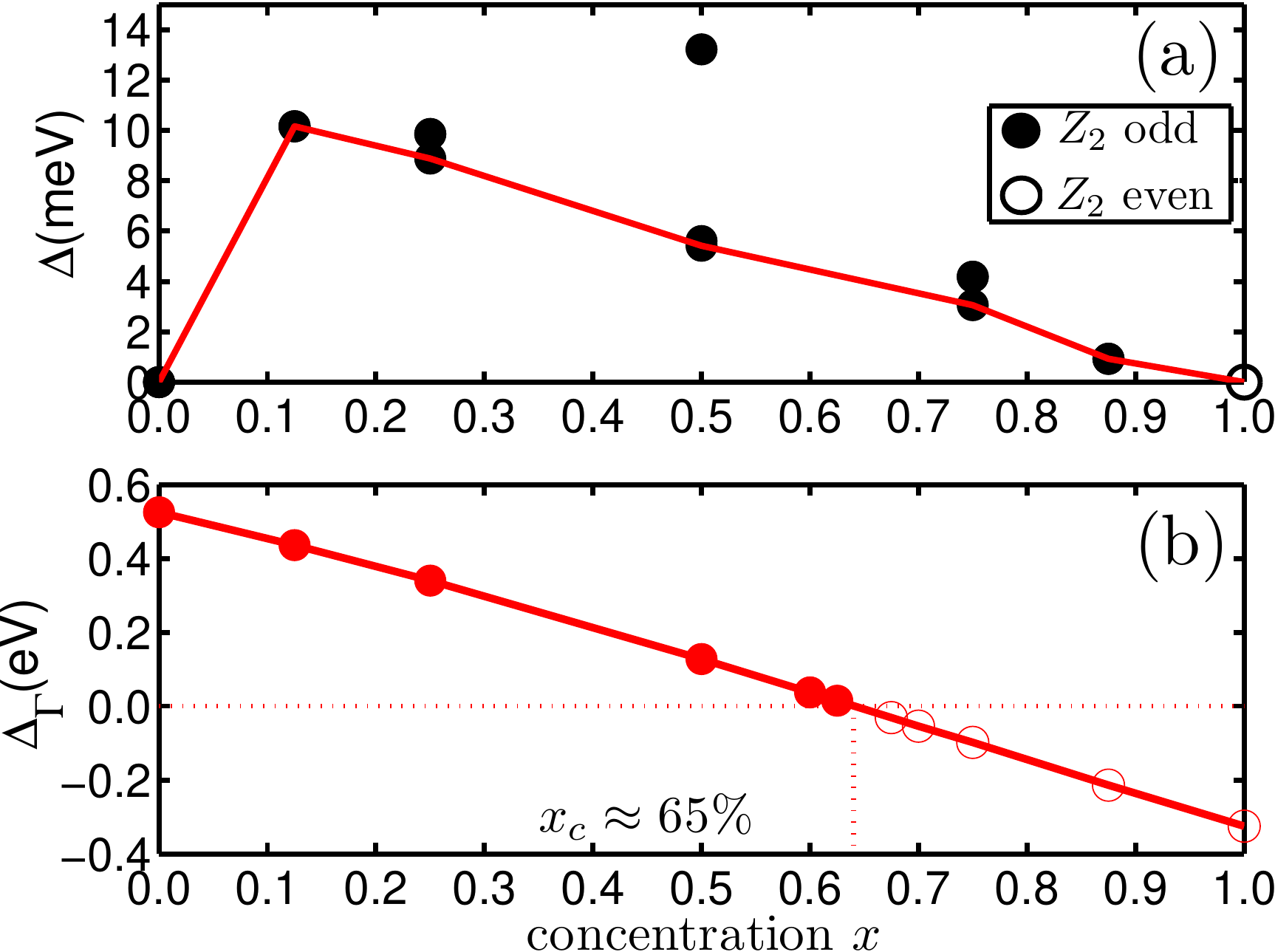}
\caption{(a) Ground-state energies vs.~impurity
concentration $x$ for \bisb\ supercells, following the same
conventions as in Fig.~\ref{fig:In-supercell}.
(b) Band gap at the center of the Brillouin zone vs.~impurity
concentration $x$ computed within the virtual crystal approximation.
Positive and negative values of band gap denote the
topological and normal phases respectively.}
\label{fig:Sb-supercell}
\end{figure}

Turning now to the case of Sb substitution, we find a quite
different behavior.  The corresponding total-energy results for
the 2$\times$2$\times$1 \bisb\ supercells are presented in
Fig.~\ref{fig:Sb-supercell}(a).  Here we find that
the energies of different configurations at the
same $x$ differ by no more than 10\,meV per primitive unit cell,
which is roughly ten times smaller than in \biin\ (note
the difference in the vertical scales here compared to
Fig.~\ref{fig:In-supercell}).  This
signifies that the disorder effect is very weak in this system.
It is also evident from Fig.~\ref{fig:Sb-supercell}(a)
that the system remains in the TI phase even up to $x\!=\!87.5\%$, in
sharp contrast to the behavior in \biin.

Because we find the disorder effect to be so weak in \sbse,
we have also analyzed its behavior using the virtual
crystal approximation (VCA), in which each Bi or Sb is replaced by
an identical average atom whose properties are a weighted mean of
the two constituents.  We implement the VCA in a Wannier basis
by constructing separate 30-band models for \bise\ and \sbse,
including all the valence cation and anion $p$ orbitals. The
Hamiltonian matrix elements
$H_{mn}^\textrm{VCA}$ of the ``virtual crystal''
are taken as the linear interpolation in $x$ of the two bulk
materials,
$H_{mn}^\textrm{VCA}=(1-x)H_{mn}^\textrm{Bi}+xH_{mn}^\textrm{Sb}$,
where $H_{mn}^\textrm{Bi}$ and $H_{mn}^\textrm{Sb}$
denote the matrix elements of the TB models of \bise\ and \sbse.
We note in passing that one has to be cautious when generating the WFs
for the VCA procedure, since it is important for the Wannier basis
functions to be as similar as possible before the averaging
takes place.  Only in this way will the addition and subtraction
between two different Hamiltonians be well defined. Because
the maximal localization procedure might generate different WFs for
different systems as it seeks to minimize the ``spread
functional,''\cite{MLWF-1} we construct the WFs for the VCA
treatment simply by projecting the Bloch states onto the same set
of atomic-like trial orbitals without any further iterative
localization procedure.

Within this VCA approach, it is straightforward to compute the band
gaps and topological indices, since only a primitive bulk cell
is needed.  Fig.~\ref{fig:Sb-supercell}(b)
shows how $\Delta_{\Gamma}$, the band gap at the Brillouin zone
center, evolves with $x$ for the \bisb\ virtual crystal.
It is evident that the gap closes at $x_c\simeq65\%$,
where the system undergoes a transition to the normal-insulator
state (here indicated by a negative gap value).

\subsection{Orbital character}
\label{sec:dos}

\begin{figure}
\centering
\includegraphics[width=8.0cm]{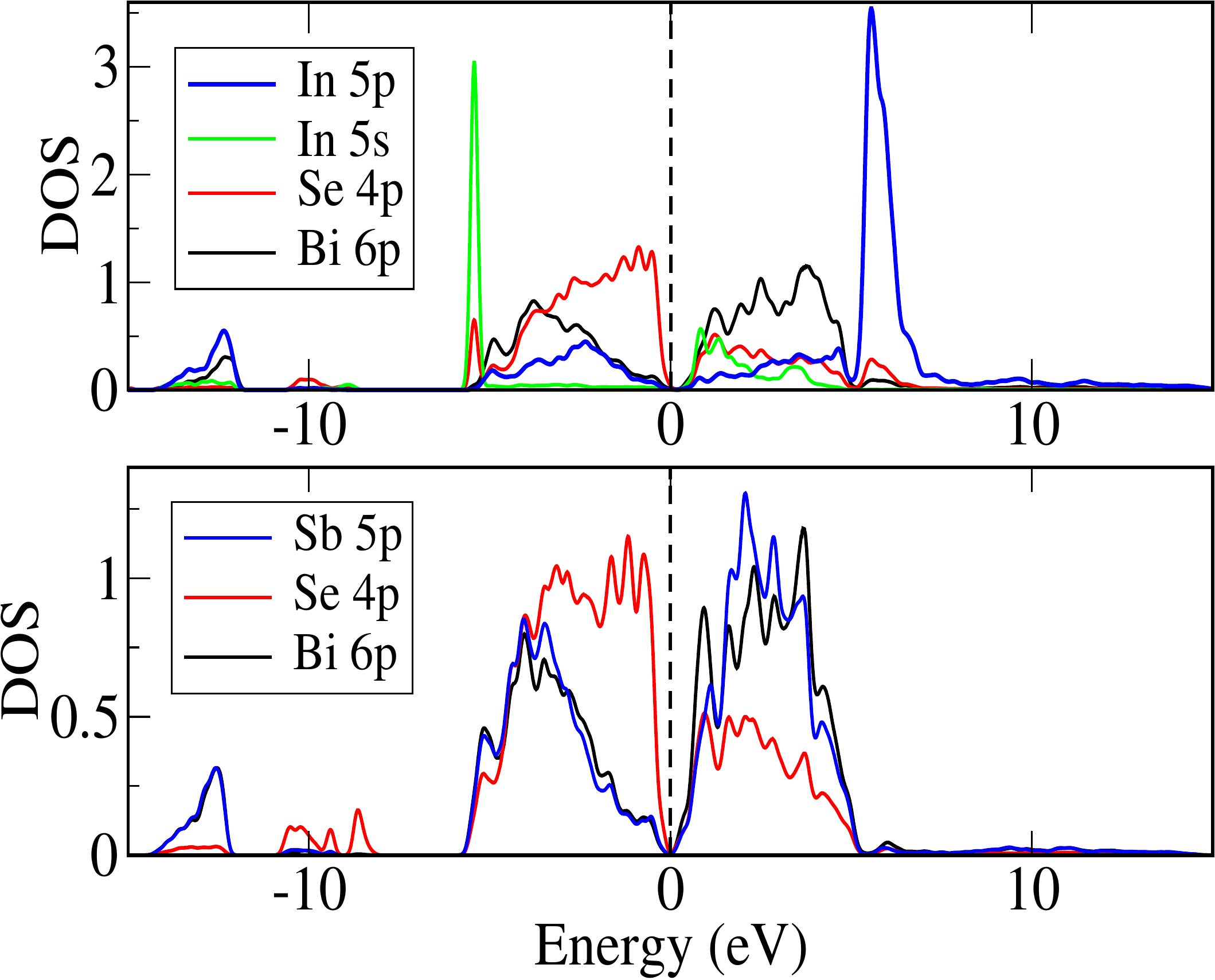}
\caption{(a) Local DOS of \biin\ at $x\!=\!12.5$\% for the $s$ and $p$ orbitals
on the substituted In atom and the $p$ orbitals on first-neighbor 
Bi and Se atoms.
(b) Local DOS of \bisb\ at $x\!=\!12.5$\% for the $p$ orbitals on the substituted Sb
atom and the $p$ orbitals on first-neighbor Bi and Se atoms.}
\label{fig:local-dos}
\end{figure}

To get some physical insight about the distinct behaviors in the
two substituted systems, we turn to study the orbital character
at a low composition of $x\!=\!12.5\%$. The local density of states
(DOS) of the substituted In and Sb atoms and their neighboring
Bi and Se atoms are plotted in Fig.~\ref{fig:local-dos}.
For low-composition In-substituted systems, the In $5s$ orbitals
and nearest-neighbor Se $4p$ orbitals form bonding and antibonding
states, with the former
leading to a flat band deep in the valence bands corresponding
to the In 5$s$ peak around $-$6\,eV
in Fig.~\ref{fig:local-dos}(a). The
hybridized $s\textrm{-}p$ anti-bonding
states further interact with the Bi $6p$
orbitals, bringing some In 5$s$ character into the conduction
bands.
The In 5$p$ orbitals
are mainly responsible for the sharp peak about 7\,eV above the Fermi level
in Fig.~\ref{fig:local-dos}(a), but also mix with Bi and Se $p$ orbitals on
the nearby atoms to contribute to the lower conduction-band states.
The hopping between In 5$p$ and neighboring Se 4$p$ states,
on the other hand, contributes mainly to the valence band, but also
to the lower conduction bands.

If one only focuses on the low-energy physics,
say within 5\,eV of the Fermi level, one would notice that the
In 5$p$ states are homogeneously distributed among the valence
and conduction bands. On the other hand, the $s$ orbitals are more
concentrated at the bottom of the valence and conduction bands. This
implies that the effects of In $5s$ and $5p$ orbitals in the
supercell electronic structure are distinct. The non-homogeneously
distributed In 5$s$ states may be crucial in determining the
topological properties of the supercell.  From the DOS at the
$\Gamma$ point (not shown here), we also observe that the VBM
is mostly
composed of Se $4p$ states, while the CBM is dominated by Bi
$6p$ states.  This implies that the nontrivial topological band inversion
has already been removed at 12.5\% of In substitution.

For the Sb substitution at $x\!=\!12.5\%$,
however, the local DOS shown in Fig.~\ref{fig:local-dos}(b)
indicates that that Sb $5p$ orbitals are more or less homogeneously
distributed among the valence and conduction bands as they hybridize
with the Bi and Se $p$ states.  In fact, the Sb 5p and Bi 6p local
DOS profiles are strikingly similar.
While not displayed,
we also explore the DOS of \bisb\ at other compositions,
and observe that the hybridization between Bi, Se and Sb $p$
states remains homogeneous over the entire composition range. A
homogeneous hybridization of Bi, Se and Sb $p$ states tends to confirm
the appropriateness of the use of VCA with artificial
orbitals to construct an effective description of the electronic structure of the
substituted system. Within the VCA, the strength of the effective SOC would
be expected to decrease linearly as $x$ is increased, suggesting that the
topological phase transition in the Sb-substituted system should
belong to the linear band-closure regime.

To study In-substituted \bise\ at a lower concentration,
a 2$\times$2$\times$2 supercell has been constructed, in which
one out of 16 Bi atoms is substituted by In.
The supercell lattice vectors are shown in Fig.~\ref{fig:latt}.
The energy cutoff is taken to be the same
as for the 2$\times$2$\times$1 supercell calculations.
A 3$\times$3$\times$3 Monkhorst-Pack $\mathbf{k}$ mesh is used
for ionic relaxation and calculation of the ground-state energy, while
it is increased to 4$\times$4$\times$4 for the non-self-consistent
calculation used to interface with Wannier90.
The ground-state energy is indicated by the filled circle at $x\!=\!6.25\%$
in Fig.~\ref{fig:In-supercell}, and is confirmed to be in the \Z2-odd\
phase from the \Z2\ index and surface-state calculations.

\subsection{Shift of In 5\textit{s} levels}
\label{sec:s-shift}

\begin{figure}
\centering
\subfigure{
\includegraphics[height=3.8cm,width=3.97cm]{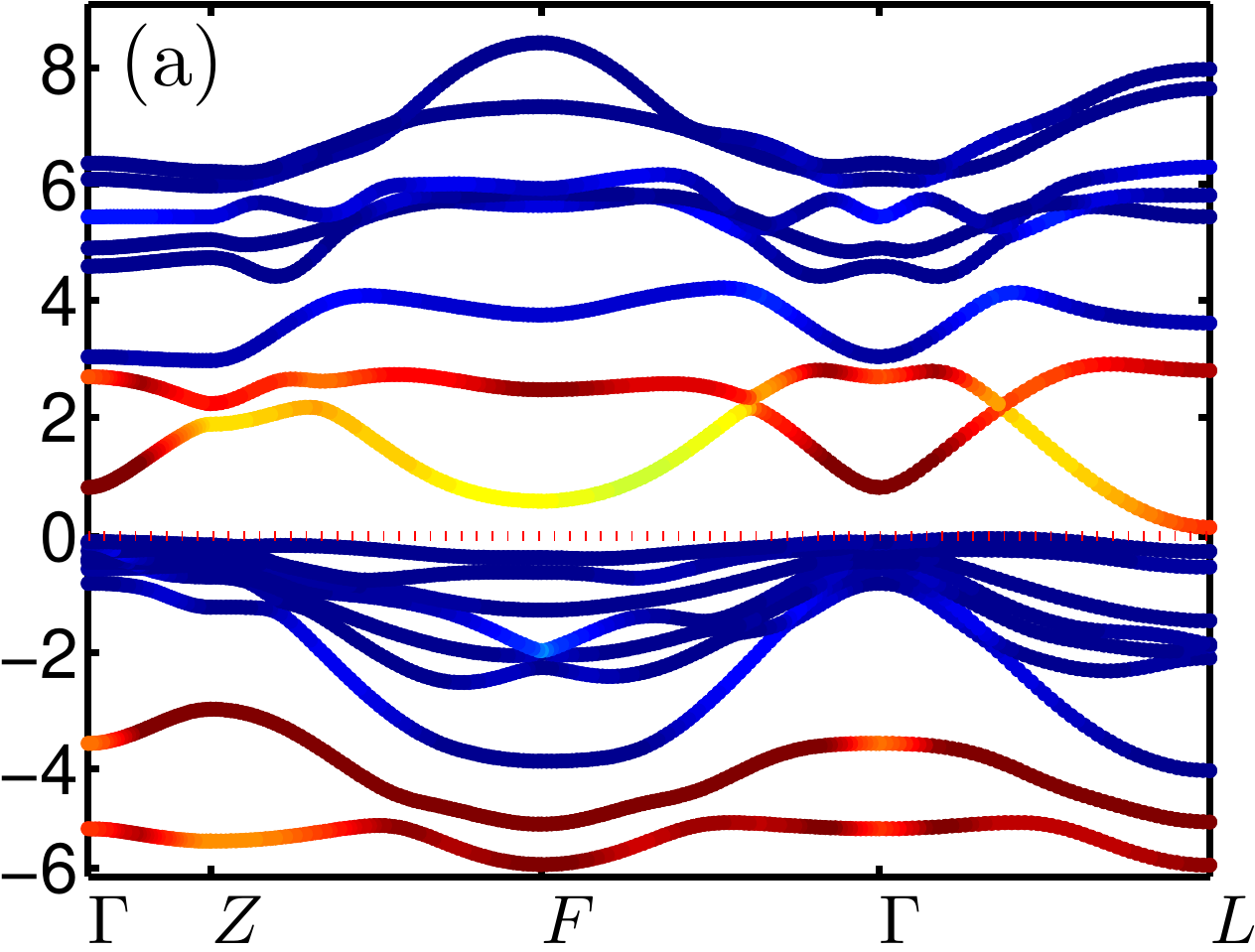}}
\hspace{0cm}
\subfigure{
\includegraphics[height=3.86cm,width=4.37cm]{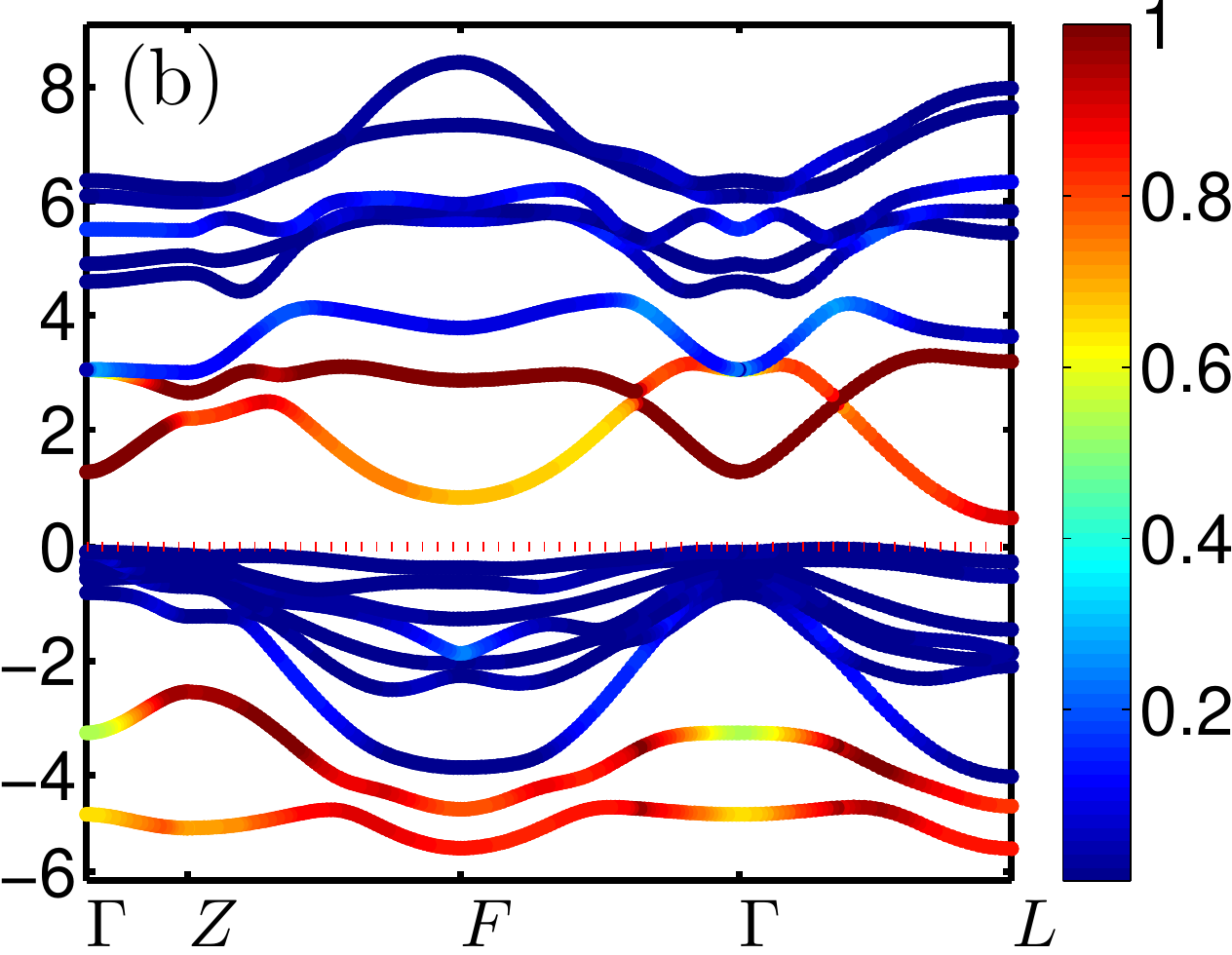}}
\caption{(a) Wannier-interpolated bandstructure of \inse,
with color code indicating In $5s$ character.
(b) Same but with In $5s$ levels shifted upward by 0.79\,eV.}
\label{fig:inse}
\end{figure}

Among the In-substituted configurations, our calculations
find that C$_{0.5}$ and C$_{0.75}$ are metallic, in contrast
with experimental observations showing the gap opening
with increasing $x$ beyond the transition to the normal
phase.\cite{oh-prl12,armitage1} The reason for the gap closure
becomes clear from an inspection of our calculated bandstructure of bulk
In$_2$Se$_3$, shown in Fig.~\ref{fig:inse}(a), which was computed
using the Wannier interpolation capabilities of the Wannier90
package\cite{wannier90} based on a 34-band TB model including
the 30 valence $p$ orbitals and four In $5s$ orbitals.  The
color coding in Fig.~\ref{fig:inse}(a) shows the degree of
In $5s$ character.  We find that there is almost a gap
closure, $E_\textrm{g}\simeq0.15$\,eV, much smaller than the
experimental value of 1.34\,eV.\cite{beta-inse}  Our small gap
clearly results from a low-lying conduction band at $L$ that is
dominantly of In $5s$ character.  For the C$_{0.5}$ and C$_{0.75}$
cases, these states get folded and mixed with other
conduction-band states in such a way as to cause the metallic
behavior observed in our supercell calculations.

We have good reason to believe, however, that the energy position
of these In $5s$ is incorrectly predicted by standard
density functional theory (DFT).\cite{hohenberg-kohn,kohn-sham}
It is well known that DFT tends to underestimate gaps,
especially when the character of the VBM
(here $p$ states at $\Gamma$) and the CBM
(here In $5s$ states at $L$) are different.  More specifically,
however, quasiparticle calculations on InAs have shown
that the In $5s$ energy levels are too low relative to the
many-body GW calculation.\cite{louie-prb91}
In particular, the CBM at $\Gamma$, having In $5s$ character, was
found there to be too low by about 0.79\,eV within DFT.
We have checked that our In $5s$ energy positions do
not depend sensitively on the use of the local-density
approximation (LDA)\cite{zunger-prb81} \textit{vs.} GGA,
the choice of pseudopotentials, or the use of different code
packages.\cite{wien2k-web,vasp}  Therefore, we conclude that
more advanced approaches such as hybrid functionals or
direct many-body methods are needed to fix this problem.

Unfortunately, application of hybrid functionals to our supercell
calculations would be computationally expensive.
Here we have taken a simpler approach to adjust the
In $5s$ energy levels.  The Wannier interpolation procedure has already
provided us with a first-principles effective TB model reproducing the
DFT bandstructure.  We simply shift the energies of all the In
$5s$ orbitals within this effective model upward by 0.79\,eV,
the value taken from Refs.~\onlinecite{louie-prb91},
and leave all other matrix elements unchanged.  The resulting
bandstructure for bulk In$_2$Se$_3$ is shown in Fig.~\ref{fig:inse}(b).
We find that the band gap opens up to 0.52\,eV, while otherwise
the general character of bandstructure is not significantly changed.

While 0.52\,eV is still far from an experimentally correct estimate of
the gap, we expect our modified Wannier Hamiltonian
should be good enough for the purpose of computing topological
properties of \biin\ solid solutions.
Once we apply this shift, we find that that the supercells that
were metallic before are now insulating, and moreover the states
near the Fermi energy that determine the topological character do not
have significant In $5s$ character.  Therefore, the magnitude of the
shift is not important for computing the topological properties,
as long as it is large enough to prevent the In $5s$ levels from
interfering.
In any case, since the $\beta$ phase of In$_2$Se$_3$ is not very
stable at room temperature (it has to be stabilized by doping
small amounts of Sb),\cite{beta-inse}
a direct comparison between the experimental and theoretical band gaps
is not very meaningful. Therefore, we adopt the procedure here of
applying the 0.79\,eV shift of In $5s$
levels in all of our In-substituted supercell calculations.
In particular, the \Z2\ indices (filled \textit{vs.} open
circles) shown in Fig.~\ref{fig:In-supercell} have been computed
using this shift, as will be discussed in detail next.

\subsection{\Z2\ indices}
\label{sec:Z2}

The strong \Z2\ indices of all the In and Sb-substituted \bise\
supercells have been calculated in order to locate the critical
concentrations for the transition from topological to normal behavior
in the two solid-solution systems.
As discussed above, some of the supercells (C$_{0.25}$, C$_{0.5}$ and
C$_{0.75}$) have inversion symmetry, in which case the strong
\Z2\ index can
be evaluated simply by counting the parities of the occupied bands at the
TRIM in the BZ. Specifically, if one defines $\delta_i$ as the
product of the parities of the occupied bands (counting just one band
from each Kramers doublet) at the $i$th TRIM in the BZ,
the strong \Z2\ index is just $\nu_0=\prod_{i=1}^{8}\delta_i$,
i.e., the product of $\delta_i$ at all the eight TRIM.\cite{ti_inversion}

In the general case, however, the strong \Z2\ index
has to be determined by explicitly calculating the
2D \Z2\ indices of the top and bottom slices of half of the 3D BZ.
There are six such 2D indices,
namely $\nu_j\equiv\nu_{k_j\!=\!0}$ and
$\nu'_{j}\equiv\nu_{k_j\!=\!\pi}$, corresponding to the indices of the
slices at $k_j\!=\!0$ and $k_j\!=\!\pi$, where $j\!=\!\{1,2,3\}$
labels the three wavevector directions in the BZ.
However, only four of the six indices turn out to be
independent variables.
The indices $\nu_1$,
$\nu_2$ and $\nu_3$
are usually taken to define the three weak topological indices,
while the product $\nu_0=\nu_{j}\nu'_{j}$
of the two indices on any pair of parallel slices is
known as the strong \Z2\ index $\nu_0$.
This means that if two parallel slices have different \Z2\ indices,
as for example at $k_3\!=\!0$ and $k_3\!=\!\pi$,
then $\nu_0$ is odd and the system is a strong TI; otherwise it is
a weak TI if any indices are odd, or normal if not.

In the absence of inversion symmetry,
the 2D \Z2\ index is defined by the change of 1D
``time-reversal polarization,'' say in the $k_1$ direction,
as the other wavevector $k_2$ evolves from $0$ to $\pi$.
The time-reversal polarization can be explicitly visualized by
tracing the 1D hybrid Wannier charge centers (WCCs)\cite{MLWF-1} in the
$k_1$ direction as a function of $k_2$.
The \Z2\ index is odd if the hybrid WCCs of the Kramers doublets switch
partners during the evolution, and even otherwise.\cite{fu-prb06}

We implement these ideas in practice using the approach of
Soluyanov and Vanderbilt,\cite{soluyanov-prb11} in which the
2D \Z2\ index is obtained by counting the number of
jumps of the ``biggest gap'' among the 1D hybrid WCCs during
the evolution.
The approach for computing the \Z2\ indices described above has been
implemented in the Wannier basis using the matrix elements of the effective
TB model and the WCCs generated from Wannier90.\cite{wannier90}

The results are shown in Figs.~\ref{fig:In-supercell} and
\ref{fig:Sb-supercell}(a) by using filled circles to indicate cases
in which the strong Z2\ index is odd, while an open circle means
it is even.  In fact, none of the \Z2-even\ configurations are found
to be weak TIs, so open circles denote topologically normal insulators.
If one follows the solid (red) lines in Fig.~\ref{fig:In-supercell} and
Fig.~\ref{fig:Sb-supercell}(a), which track the configurations
with lowest energies, it is clear that for
\biin\ the system becomes topologically
trivial for $x\!>\!6.25 \%$. For \bisb, however,
the TI phase is preserved up to 87.5$\%$.

It should be emphasized again that
a 0.79\,eV shift has been added on the on-site In $5s$ energy levels
in the effective TB models for the supercells of \biin.
However, except for C$_{0.5}$ and C$_{0.75}$, which are
metallic without the shift, the \Z2\ indices of all the other
configurations are unchanged by the application of this shift.

\subsection{Effects of In $5s$ orbitals on bulk and surface states}
\label{sec:surface}

To understand why the phase transition happens so rapidly in \biin,
we focus on the $x\!=\!0.125$ supercell, and separately investigate
the effects of In $5s$ orbitals and SOC on bulk bandstructure.
As shown in Table~\ref{table:In-5s}, without
SOC the In $5s$ orbitals try to pull down the VBM, leading to a
band gap as large as 0.7\,eV at $\Gamma$, such that the SOC strength
is not large enough to invert the CBM and VBM. If the In $5s$ orbitals are
removed, however, the gap at $\Gamma$ is only 0.42\,eV without SOC,
and when SOC is added back the band inversion reoccurs,
with an inverted gap as large as 0.26\,eV (denoted with a minus sign in
Table~\ref{table:In-5s}). We also notice that the
shift of In $5s$ levels only changes the gap at $\Gamma$ by 0.04\,eV,
and does not influence the topological behavior.

\begin{table}
\caption{\label{table:In-5s}
Bulk band gaps at $\Gamma$ for $x\!=\!12.5\%$ in \biin.
``Shifted'' In $5s$ levels were raised by 0.79\,eV (see text).}
\centering
\begin{tabular}{lcc}
\hline\hline
  & without SOC &  with SOC \\
  & (eV) & (eV) \\
\hline
In $5s$ levels unshifted & 0.72   & 0.11 \\
In $5s$ levels shifted & 0.68 & 0.07 \\
In $5s$ levels removed & 0.42  & \hspace{-8pt}$-$0.26 \\
\hline
\end{tabular}
\end{table}

\begin{figure}
\includegraphics[width=8.6cm]{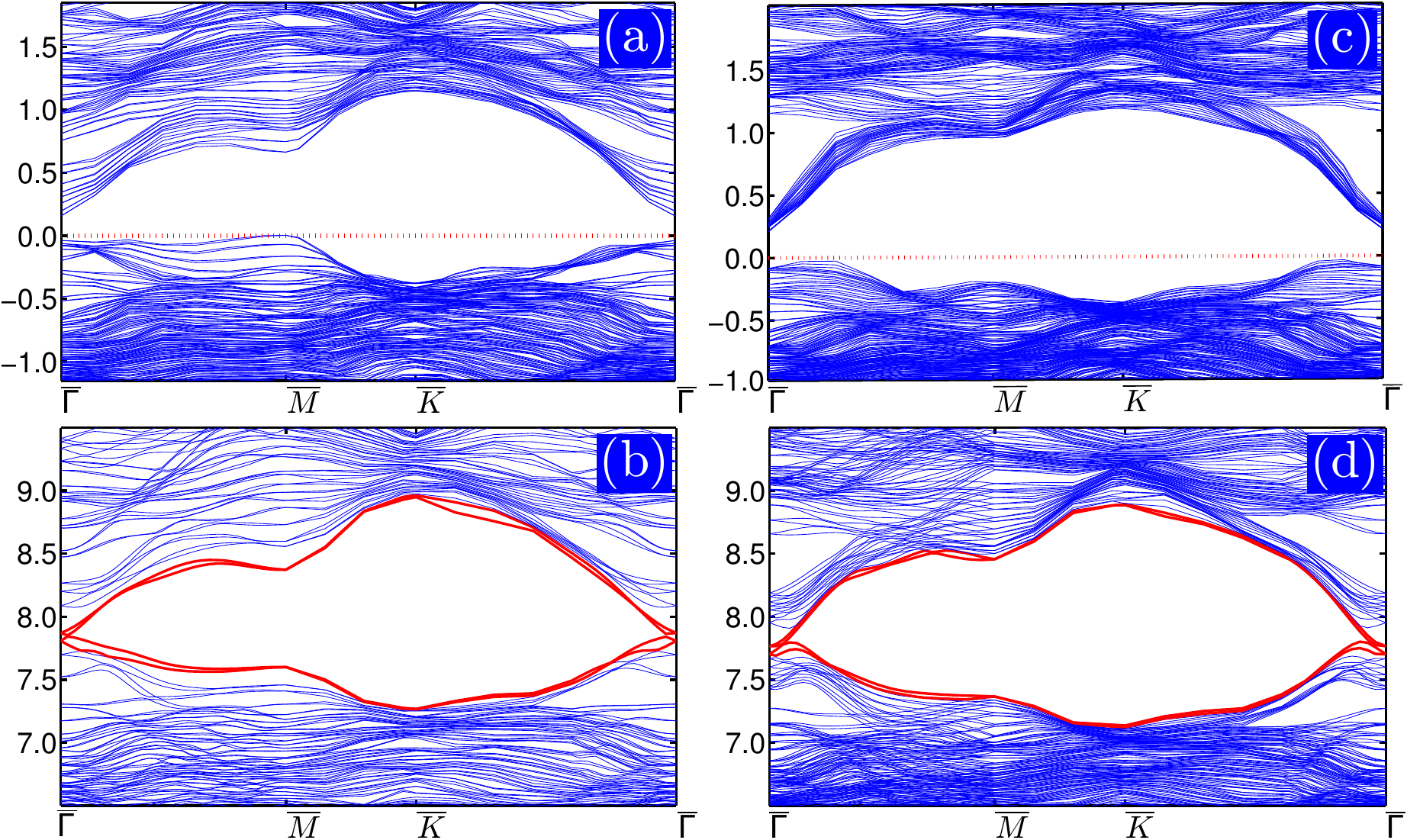}
\caption{
Surface bandstructures for (Bi$_{1-x}$In$_{x})_2$Se$_3$ slabs,
plotted in the 2D surface Brillouin zone.  (Surface states are shown in red.)
(a) 8QL slab for $x\!=\!0.125$.
(b) 4QL slab for $x\!=\!0.125$ but with In $5s$ orbitals removed.
(c) 12QL slab for energetically favored configuration
C$'_{0.25}$ at $x\!=\!0.25$.
(d) 8QL slab for C$'_{0.25}$ with In $5s$ orbitals removed.
Note split Dirac cones arising at $\overline{\Gamma}$ in
(b) and (d).}
\label{fig:ss}
\end{figure}

We continue to study the effects of In $5s$ orbitals on surface states
by calculating the surface bandstructures both with and without In $5s$
orbitals. The surface bandstructures shown in Fig.~\ref{fig:ss}
are calculated with the ``slab method,''
where the first-principles TB models of slabs of In- and Sb-substituted
\bise\ with finite thickness stacked along
the [111] direction have been constructed. This is done
by extrapolating the matrix elements of the primitive unit cell TB
model to multiple QLs along the [111] direction, then truncating
at the two surfaces to enforce open boundary conditions.
The 2D surface bandstructure is then
obtained by directly diagonalizing the TB model of the slab.

It has to be noted that the surface states are not calculated
self-consistently by doing such a truncation at the surface,
because the Wannier functions close to the surface could be
significantly deformed and the hopping parameters between orbitals
close to the surface are expected to be different
from those deep in the bulk. These effects
are not properly included simply by truncating at
the slab boundaries.  However, we argue that even though these
surface effects could be important in determining such details
as the exact position of the Dirac point (if present) relative
to the bulk CBM, they cannot change the topological character of
the surface states, which is what we really care about here.

To understand
the role of In $5s$ orbitals in the phase-transition process,
the surface states are calculated both with and without the
In $5s$ orbitals.
The results for the configurations with lowest energy
(the In-clustered configuration) are summarized in
Table~\ref{table:surface-states},
where `$\checkmark$' indicates the existence of a Dirac cone around
$\overline{\Gamma}$, and $\times$ denotes the absence of such
a Dirac cone. The thicknesses
of the slabs are chosen such that the interference between the
states from two opposite surfaces is negligible.
\begin{table}
\caption{\label{table:surface-states}
Existence of topological surface states vs.~impurity concentration
$x$ in \biin\ and \bisb\ based on slab
calculations in the Wannier representation.}
\centering
\begin{tabular}{l c c c c c c c c}
\hline\hline
 & 0$\%$ &  6.25$\%$ & 12.5$\%$ & 25$\%$ & 50$\%$ & 75$\%$ &
      87.5$\%$  & 100$\%$ \\ [0.5ex]
\hline
With In $5s$ & $\checkmark$   & $\checkmark$ & $\times$ & $\times$ &
    $\times$ & $\times$ & $\times$ & $\times$ \\
Without In $5s$ & $\checkmark$   & $\checkmark$ & $\checkmark$ &
     $\checkmark$ & $\times$ & $\times$ & $\times$ & $\times$\\
Sb substitution& $\checkmark$  & - & $\checkmark$ & $\checkmark$ &
     $\checkmark$ & $\checkmark$ & $\checkmark$ & $\times$ \\ [1ex]
\hline
\end{tabular}
\end{table}

As can be seen in Table~\ref{table:surface-states},
the In $5s$ orbitals are directly responsible for removing the Dirac cones
from the surface spectrum for the C$_{0.125}$ and C$'_{0.25}$ cases.
This is shown explicitly in Fig.~\ref{fig:ss}, where Dirac cones
emerge at $\overline{\Gamma}$ only when the In $5s$ orbitals are removed.
(Actually, a close inspection of the figure shows a split pair of
Dirac cones contributed by opposite surfaces, where the splitting
arises because of broken inversion symmetry due to the pattern
of In substitution.)
Scanning over the Sb-substituted Bi$_2$Se$_3$ supercells from
$x\!=\!0.125$ to 0.875, clear signatures of Dirac cones are observed
for all of them (not shown here),
consistent with the results
of the bulk \Z2$\text{-}$index calculations of
Fig.~\ref{fig:Sb-supercell}(a).

Our surface-state calculations are consistent with the results
of a recent scanning tunneling microscope (STM) study of the
same system,\cite{shen} where a local suppression of density of
states was observed in the
topological surface states due to substitutional In atoms. Based
on our supercell calculations, we conclude that
this suppression of the topological surface states results from the In
$5s$ orbitals,
with an In concentration of $x\!=\!0.125$ or $x\!=\!0.25$ being sufficient
to remove them entirely.

\subsection{Disordered spectral functions}
\label{sec:spectral}

The previous superlattice calculations enabled us to capture some
important physics of the phase-transition behavior, but it
is still difficult to give a precise estimate of the critical
concentrations because of the limited size of the supercells
and the approach of studying one particular configuration at
a time.  Here we use the Wannier representation to construct
ensembles representing the disordered systems in much larger
supercells, in an attempt to study the effects of disorder in
the phase-transition process more realistically and estimate the
critical points more accurately.

Two issues arise when disorder is included. First, in a periodic
lattice structure without any disorder, the eigenstates of
are Bloch states which are perfectly coherent with
infinite lifetime, and the wavevector $\mathbf k$ is a good quantum
number. For such systems with nontrivial band topology, the easiest
way to study the topological phase transition is to look at the band
structure; a band-gap closure usually implies a phase transition
from a topological to trivial insulator. In disordered systems,
however, the ``band-gap closure'' is not so easy to recognize, because
the Bloch functions are no longer the eigenstates of the system and
a bandstructure is really not well-defined.
Secondly, as we know, the \Z2\ index is computable  for periodic lattices
if the information of occupied Bloch states in the entire BZ is
given. However, it is a difficult question how to define the \Z2\ index and
determine the topological behavior of a realistic disordered system.

Our answer to the first
issue is to look at the disorder-averaged spectral functions computed
from a large supercell with different impurity configurations,
but unfolded back to the BZ of the primitive unit cell.
If the disorder is weak, one should see a sharp spectrum with narrow
lifetime
broadening, which means the quantum states would remain coherent over
long distances. For strongly disordered systems, however, it is expected
that the spectral functions should be strongly smeared out
due to the strong randomness of the impurity scattering,
and the quantum states would be localized around the impurities
with a relatively short localization length.

We propose the ``\Z2$\text{-}$index statistics'' to address
the second problem. To be specific, several different impurity
configurations are generated in our calculations at each impurity
composition $x$, forming a representative ensemble of the disordered
system,
and we then compute the strong \Z2\ index for each configuration.
If each configuration in the disorder ensemble
is equally weighted, then when over half of the configurations are
in the \Z2-odd\ phase, we say that the disordered system can be
statistically considered as a TI.

To calculate the disordered spectral functions, we use the
Wannier effective-Hamiltonian approach as well as the technique
of unfolding first-principle supercell
bandstructures (spectral functions) as proposed by
Ku and coworkers.\cite{weiku1,weiku2}We note in passing 
that a similar technique has been proposed by Popescu and Zunger.
\cite{zunger-prl10}
To be explicit, we first construct a TB model for a 4$\times$4$\times$3
supercell of pure \bise\ whose matrix elements
are extrapolated from the primitive-cell \bise\ TB model.
The 4$\times$4$\times$3 supercell of the bulk material acts as the
reference system in which the Bi atoms are randomly substituted by
the impurity atoms. For a 4$\times$4$\times$3 supercell, there are 240
atoms of which 96 are Bi-like, so
the impurity composition $x$ can be varied on the scale of one percent,
enabling us to determine the critical point $x_c$ with
high precision.

The next procedure is to extract the Hamiltonian of a single
impurity defined under the same WF basis.
This is done by
working in a small supercell (2$\times$2$\times$1 in our case)
and subtracting the pure bulk-material Hamiltonian $H^{0}$ from the
Hamiltonian $H^s$ of a supercell containing one substituted impurity
of type $s$.  To set
the notation, we label (5-atom) cells within the supercell by $l$, sites
within the cell as $\tau$, and orbitals within the cell as $m$.
Then the impurity potential is constructed as
\begin{align}
\Delta^s_{lm,l'm'}(\tau_s)=&
(H^s_{lm,l'm'}(0\tau_s)-H^0_{lm,l'm'}) \nonumber\\
&\qquad\times\;P_{lm,l'm'}(\tau_s) \,.
\label{equa:impurity}
\end{align}
This describes the change in the on-site energy if $(lm)\!=\!(l'm')$,
or in the hopping if $(lm)\!\ne\!(l'm')$, induced by the presence
of the impurity of type $s$ on site $\tau_s$ in the central cell
$l_s\!=\!0$ of the small supercell.
We define a ``partition function''\cite{weiku2} $P_{lm,l'm'}(\tau_s)$
that is used to partition the contribution of the single impurity
from the super-images in the neighboring supercells, such that the
single impurity Hamiltonian is not influenced by the artificial
periodicity of the supercell.  In our calculations this partition
function is chosen as
\begin{equation}
P(d)= \begin{cases}
e^{-(d/r_0)^8} & \hbox{if $d\leq d_{\rm c}$} \\
0              & \hbox{otherwise}  \end{cases}
\end{equation}
where
$d=d_{lm,l'm'}(\tau_s)\equiv
   |\mathbf{r}_{lm  }-\mathbf{r}_{0\tau_s}|
 + |\mathbf{r}_{l'm'}-\mathbf{r}_{0\tau_s}|$
is chosen as a measure of the ``distance'' from the hopping matrix
element $(lm,l'm')$ to the impurity site located at
$\mathbf{r}_{0\tau_s}$.\cite{weiku2}
Here we choose $d_{\rm c}\!=\!8.69$\,\AA\ and $r_0\!=\!7.86$\,\AA. (We find
that if $d_{\rm c}>8.5$\,\AA\ and $r_0$ is chosen
slightly smaller than or equal to $d_{\rm c}$,
the impurity Hamiltonian becomes
insensitive to small variations of $d_{\rm c}$ and $r_0$.) Our partition
scheme has been tested to be able to reproduce the first-principles
2$\times$2$\times$1 supercell bandstructures at $x\!=\!0.125$ and 0.25.

We extract this impurity potential once and for all for an In atom
substituting for the top Bi atom in the quintuple-layer ($s\!=\!1$) and
again when it substitutes for the bottom Bi atom ($s\!=\!2$).
Then for a particular impurity configuration $\mathcal{R}=
\{l_1s_1,l_2s_2,\ldots\}$ of the $4\times4\times3$ supercell,
where $l_js_j$ specifies the subcell
and type of impurity and $j$ runs over the impurities in the
supercell, the effective Hamiltonian is taken as a linear
superposition of the reference-system Hamiltonian $H^{0}$
and the single-impurity Hamiltonians residing on the specified
sites in the large supercell, i.e.,
\begin{equation}
H_{lm,l'm'}^{\mathcal{R}}=H^0_{lm,l'm'}
+\sum_j \Delta^{s_j}_{(l-l_j)m,(l'-l_j)m'}(\tau_j) \,.
\label{equa:Heff}
\end{equation}

The linear superposition of the matrix
elements of different TB Hamiltonians is well-defined only when these
Hamiltonians are treated under the same WF basis. In other words,
each of the orbitals from the large supercell with impurities should
map appropriately to the corresponding orbitals of the unperturbed
reference system.  For this reason, we skip the maximal localization
procedure when generating the WFs, and instead
simply use the projection method to generate a basis that remains
in close correspondence to the atomic-like orbitals.  Once the
effective Hamiltonians has been obtained for an ensemble of
impurity configurations representing a given concentration $x$,
we calculate the spectral function for each, and unfold it
from the highly compressed supercell BZ into the primitive-cell
BZ.\cite{weiku1} Finally, the ensemble average of the unfolded
spectral functions can then be taken to reflect the effects of
disorder on the original bulk electronic states.

To be specific, let $A_{N}(\omega,\mathbf{K})$ be the spectral
function at energy $\omega$ associated with band $N$ in the supercell, with
$\mathbf{K}$ specifying the wavevector in the small supercell
BZ, given by the imaginary part of the retarded Green's function
operator $G$ via
$A\!=\!-\pi^{-1}\Im G\!=\!-\pi^{-1}\Im (\omega+i\eta-H)^{-1}$,
where $H$ is the supercell Hamiltonian and $\eta>0$ is a
small artificial smearing factor.
Then, to unfold the supercell spectral
function onto a complete set of primitive-cell Bloch states,
one can expand the primitive-cell spectral function
$A_{n}(\omega,\mathbf{k})$ in terms of the supercell spectral functions as
\begin{eqnarray}
A_n(\omega, \mathbf{k}) &=
& \sum_{N\mathbf{K}}|\langle\psi_{N\mathbf{K}}|\psi_{n\mathbf{k}}
       \rangle|^2A_N(\omega,\mathbf{K}) \,,
\label{equa:unfolding}
\end{eqnarray}
where  $|\psi_{n\mathbf{k}}\rangle$ and
$|\psi_{N\mathbf{K}}\rangle$ are the primitive-cell and supercell
Bloch states respectively, and $n$ and $\mathbf{k}$ represent
the band index and wavevector of the primitive cell.
One can solve for the coefficient
$\langle\psi_{N\mathbf{K}}|\psi_{n\mathbf{k}}\rangle$ within the
Wannier basis provided that the supercell Hamiltonian is defined
under a set of WFs having a clear one-to-one mapping with
the primitive-cell WFs by primitive-cell lattice translations,
as can be realized by using simple projection for the Wannier
construction.\cite{comment_unfolding}

\begin{figure}
\centering
\begin{minipage}[c]{0.5\textwidth}
\subfigure{\includegraphics[height=3.5cm,clip]{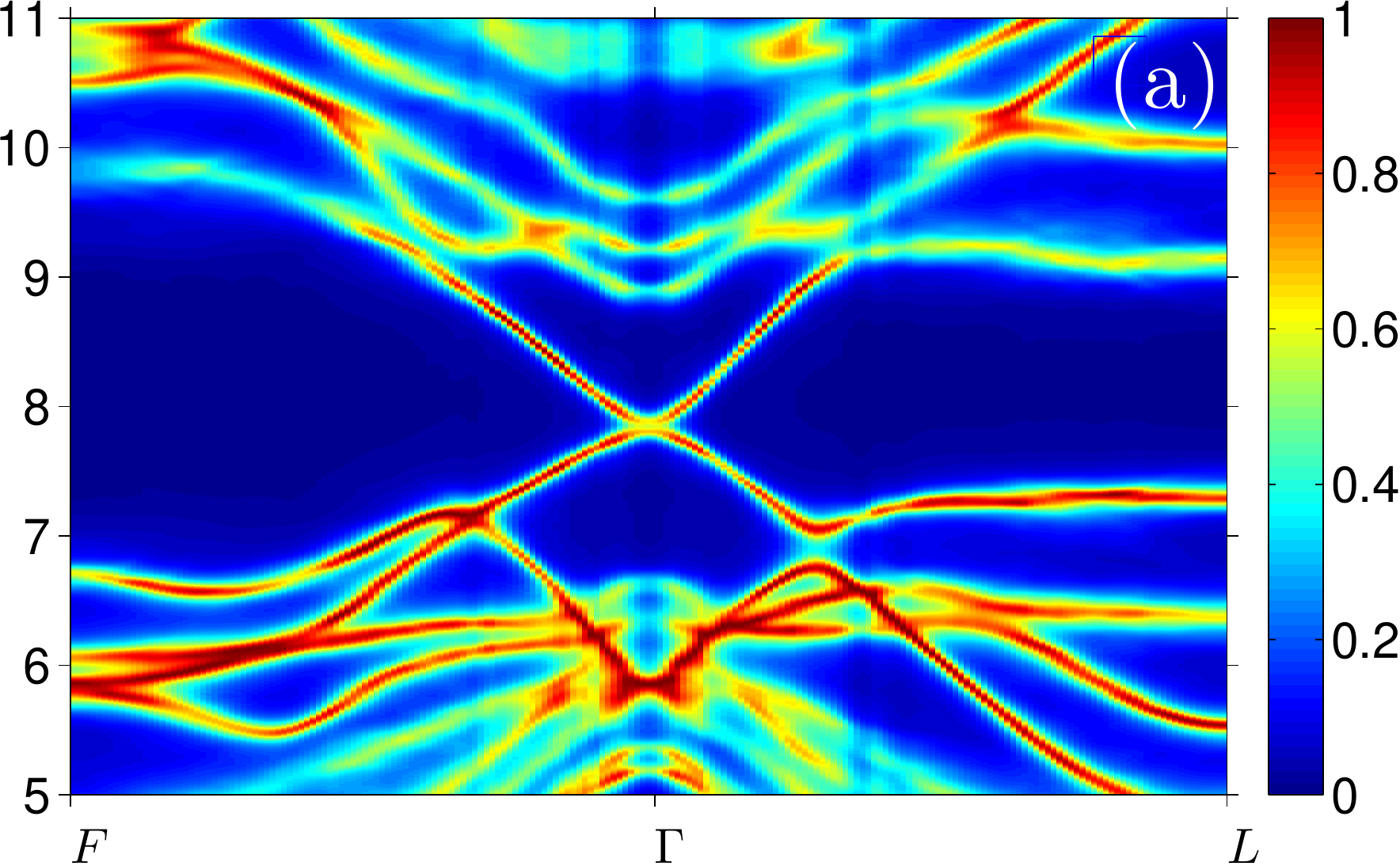}}
\end{minipage}%
\vspace{-0.1cm}
\begin{minipage}[c]{0.5\textwidth}
\subfigure{\includegraphics[height=3.5cm,clip]{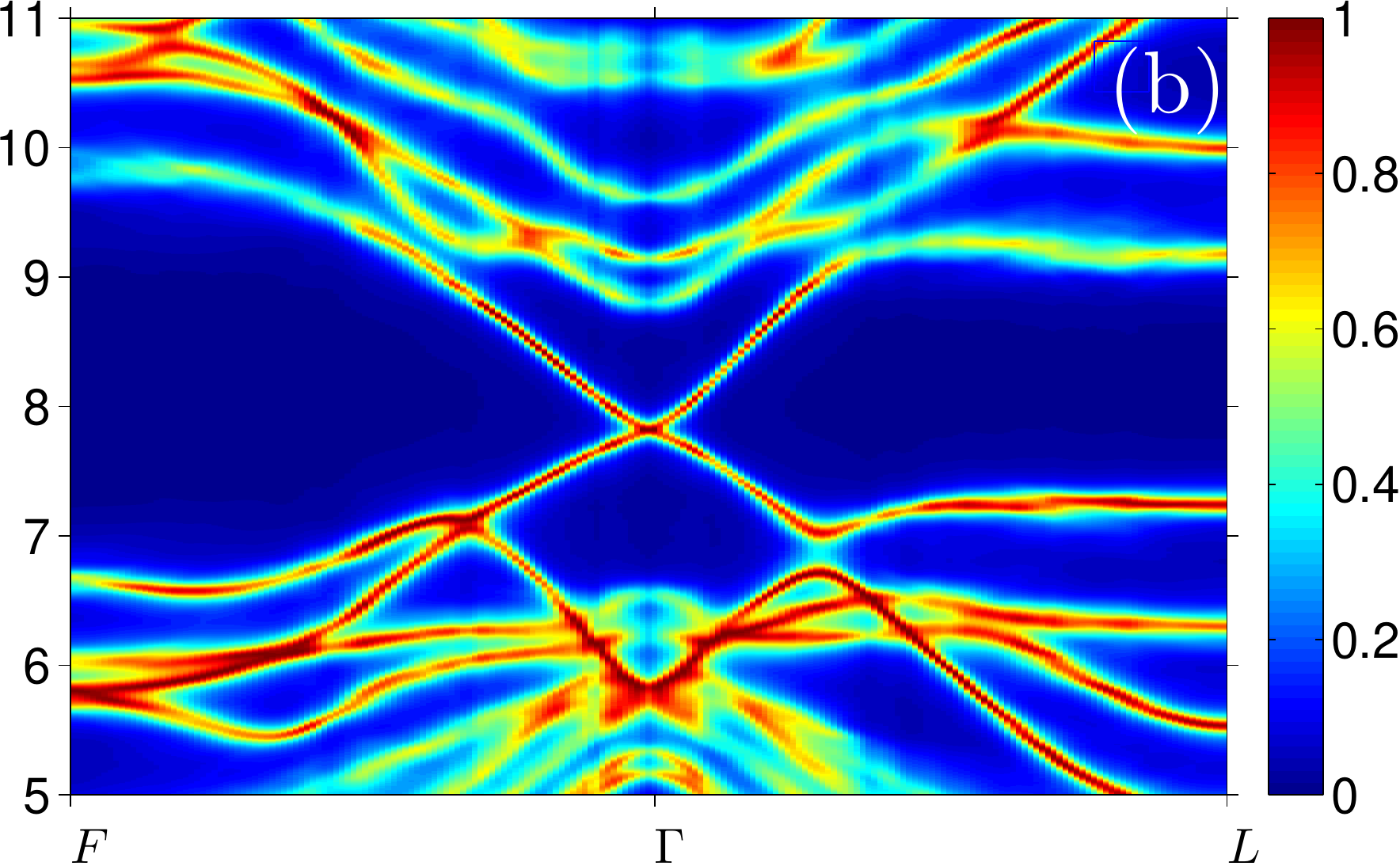}}
\end{minipage}%
\vspace{-0.1cm}
\begin{minipage}[c]{0.5\textwidth}
\subfigure{\includegraphics[height=3.5cm,clip]{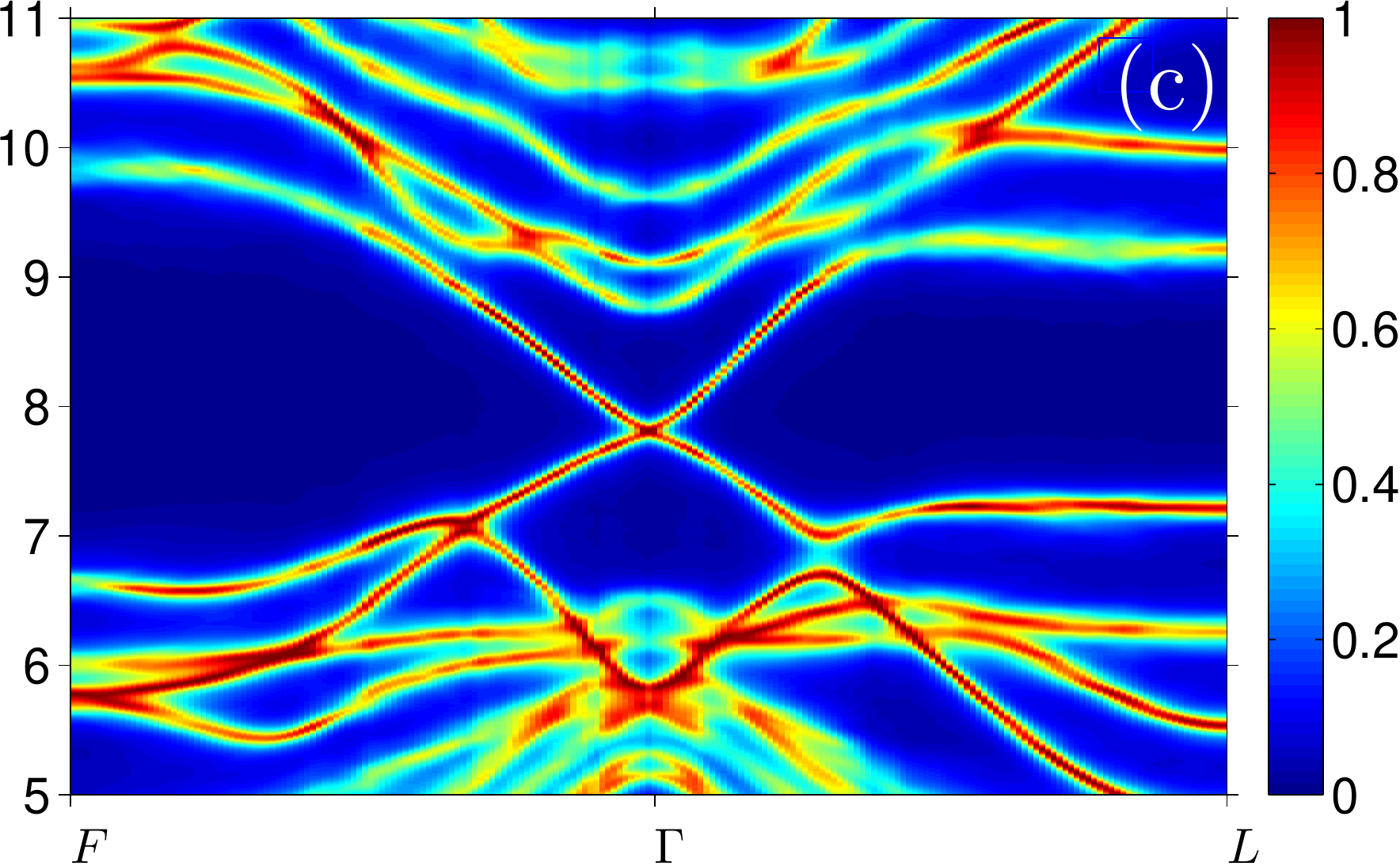}}
\end{minipage}%
\vspace{-0.1cm}
\begin{minipage}[c]{0.5\textwidth}
\subfigure{\includegraphics[height=3.5cm,clip]{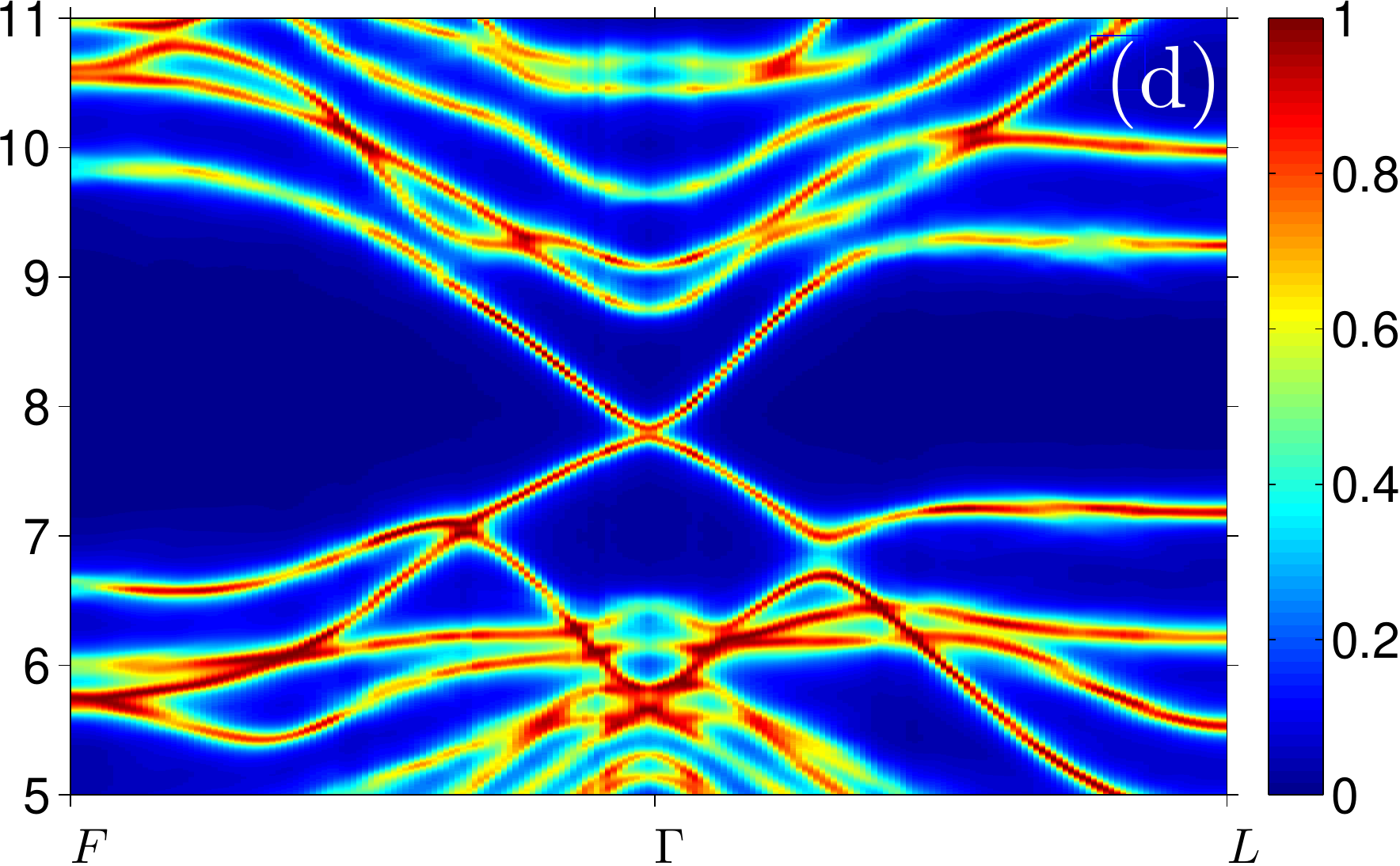}}
\end{minipage}%
\vspace{-0.1cm}
\begin{minipage}[c]{0.5\textwidth}
\subfigure{\includegraphics[height=3.5cm,clip]{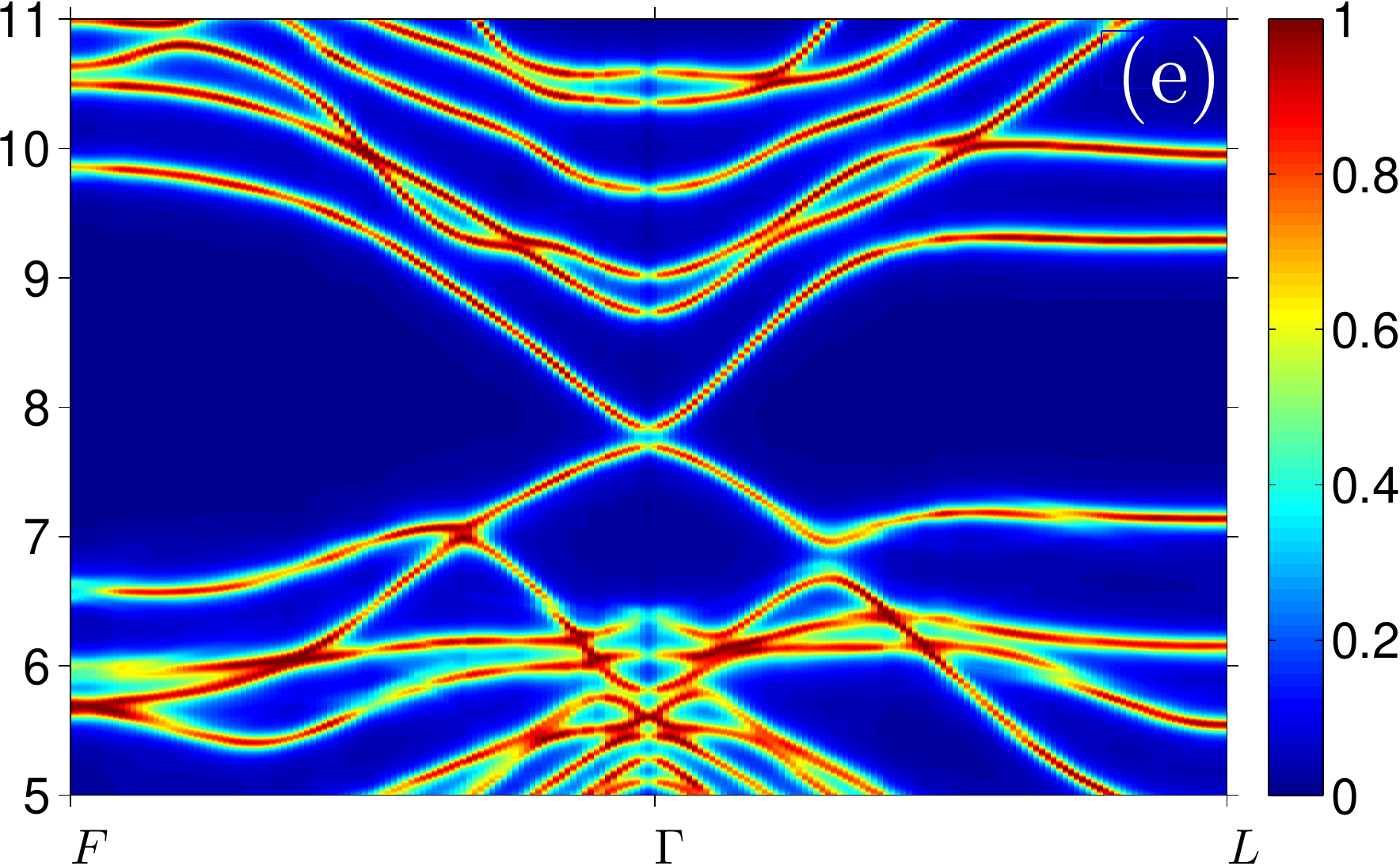}}
\end{minipage}%
\caption{Disordered spectral functions of \bisb\ 
unfolded into the primitive-cell BZ.  (a) $x\!=\!68$\%. (b) $x\!=\!78$\%.
(c) $x\!=\!83$\%. (d) $x\!=\!88.5$\%. (e) $x\!=\!99$\%.}
\label{fig:Sb-spectral}
\end{figure}

In our calculations, 16 configurations are generated for
\biin\ at each impurity composition, whereas eight configurations
are generated  for each $x$ of \bisb\
due to the much weaker effect of disorder. For \bisb\ the configurations
are generated randomly, as different configurations seem to be equally
favored energetically. For
\biin, however, the configurations are generated using the
Metropolis Monte Carlo
method according to a proper Boltzmann weight in order to
reflect the tendency of In segregation. The Boltzmann weight is
proportional to $e^{-(E_{p}N_{p})/(k_BT_g)}$,
where $E_{p}=[E(\text{C}'_{0.25})-E(\text{C}_{0.25})]/2$
is defined as the ``paring energy'' of In atoms, $N_p$ is the number of
In pairs in a particular configuration, $k_B$ is the Boltzmann constant, and
$T_g\!=\!850^\circ$C is taken as the growth temperature of the
In-substituted \bise\ sample.

\begin{figure}
\centering
\includegraphics[width=8.6cm]{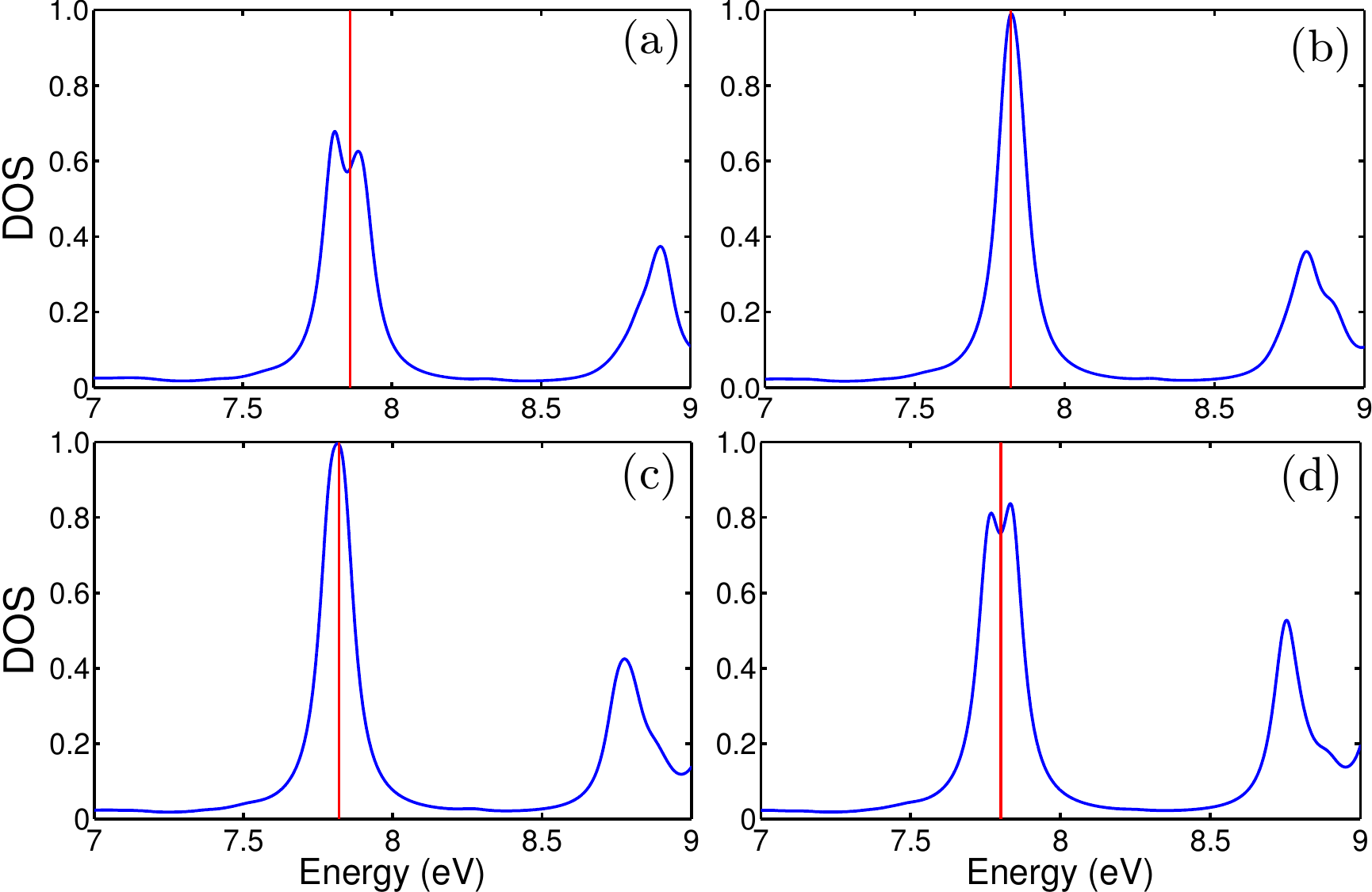}
\caption{Disordered spectral functions of \bisb\ at the $\Gamma$
point in the primitive-cell BZ for (a) $x\!=\!68$\%,
(b) $x\!=\!78$\%, (c) $x\!=\!83$\%, and (d) $x\!=\!88.5$\%.
Vertical (red) line indicates Fermi energy.
Distinct VBM and CBM peaks are still visible in the topological
phase in (a), merge in (b) and (c), and reappear in (d) as the
gap reopens in the normal phase.}
\label{fig:Sb-spectral-Gamma}
\end{figure}
\begin{figure}
\centering
\begin{minipage}[c]{0.5\textwidth}
\subfigure{\includegraphics[height=3.5cm,clip]{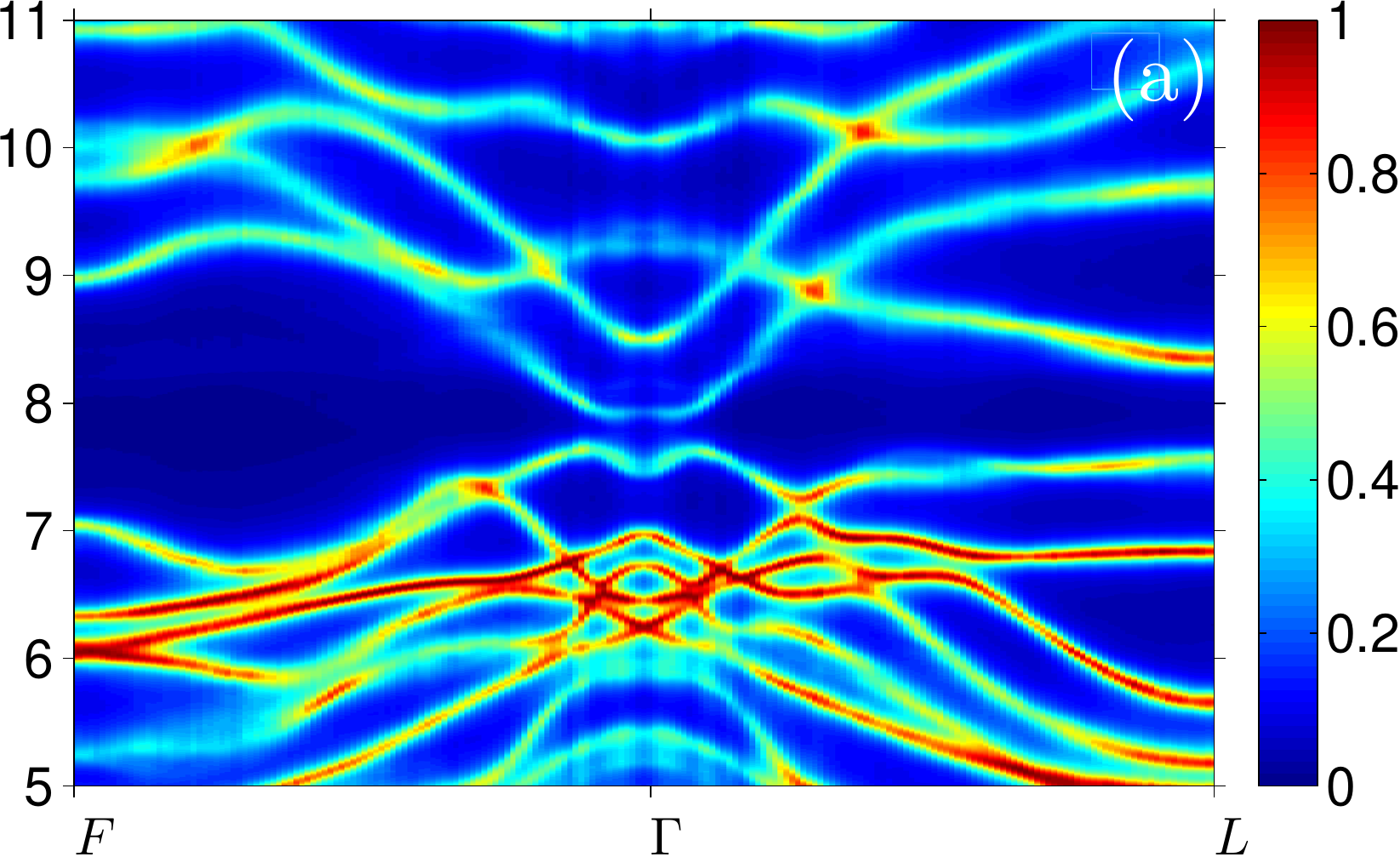}}
\end{minipage}%
\vspace{-0.1cm}
\begin{minipage}[c]{0.5\textwidth}
\subfigure{\includegraphics[height=3.5cm,clip]{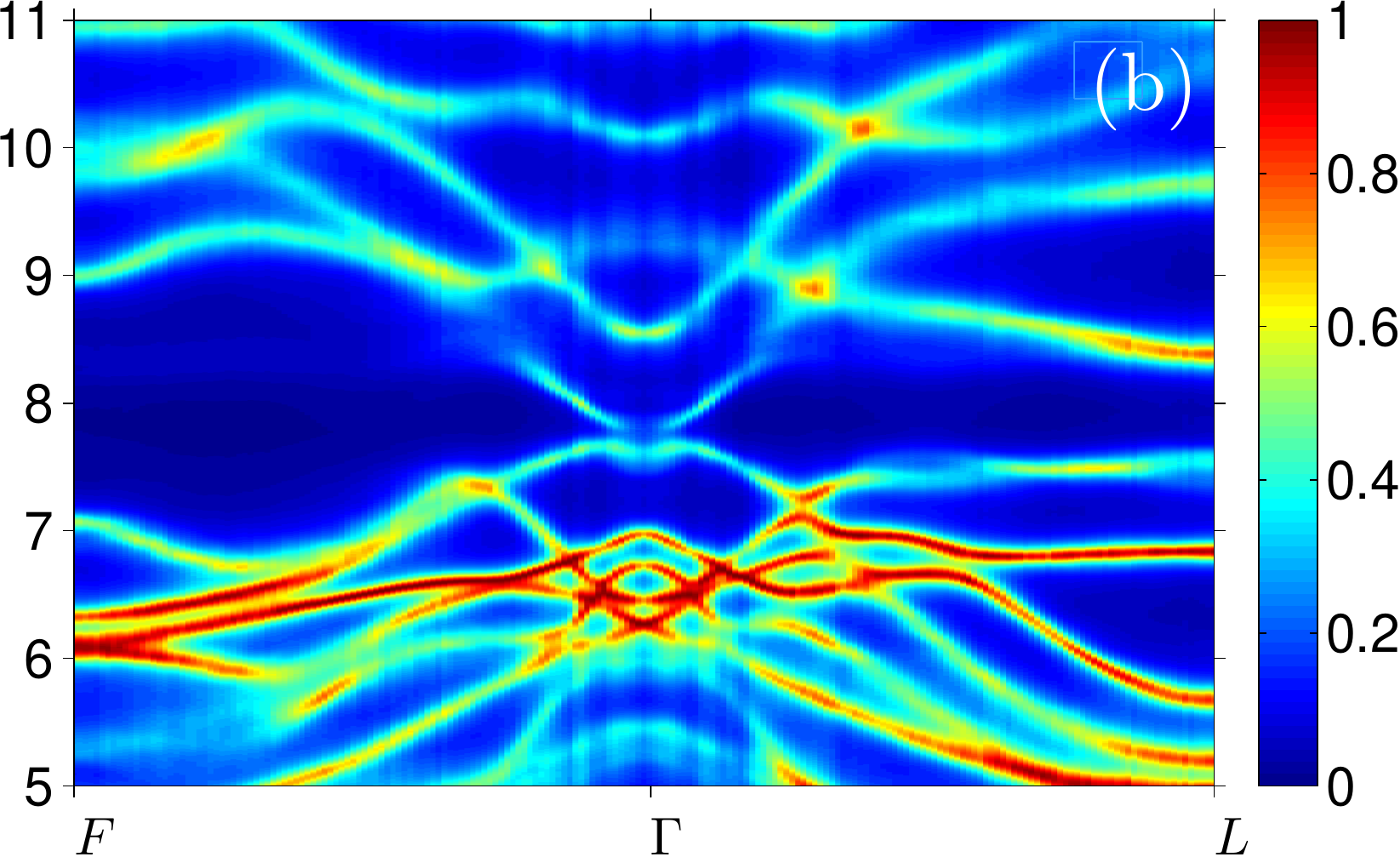}}
\end{minipage}%
\vspace{-0.1cm}
\begin{minipage}[c]{0.5\textwidth}
\subfigure{\includegraphics[height=3.5cm,clip]{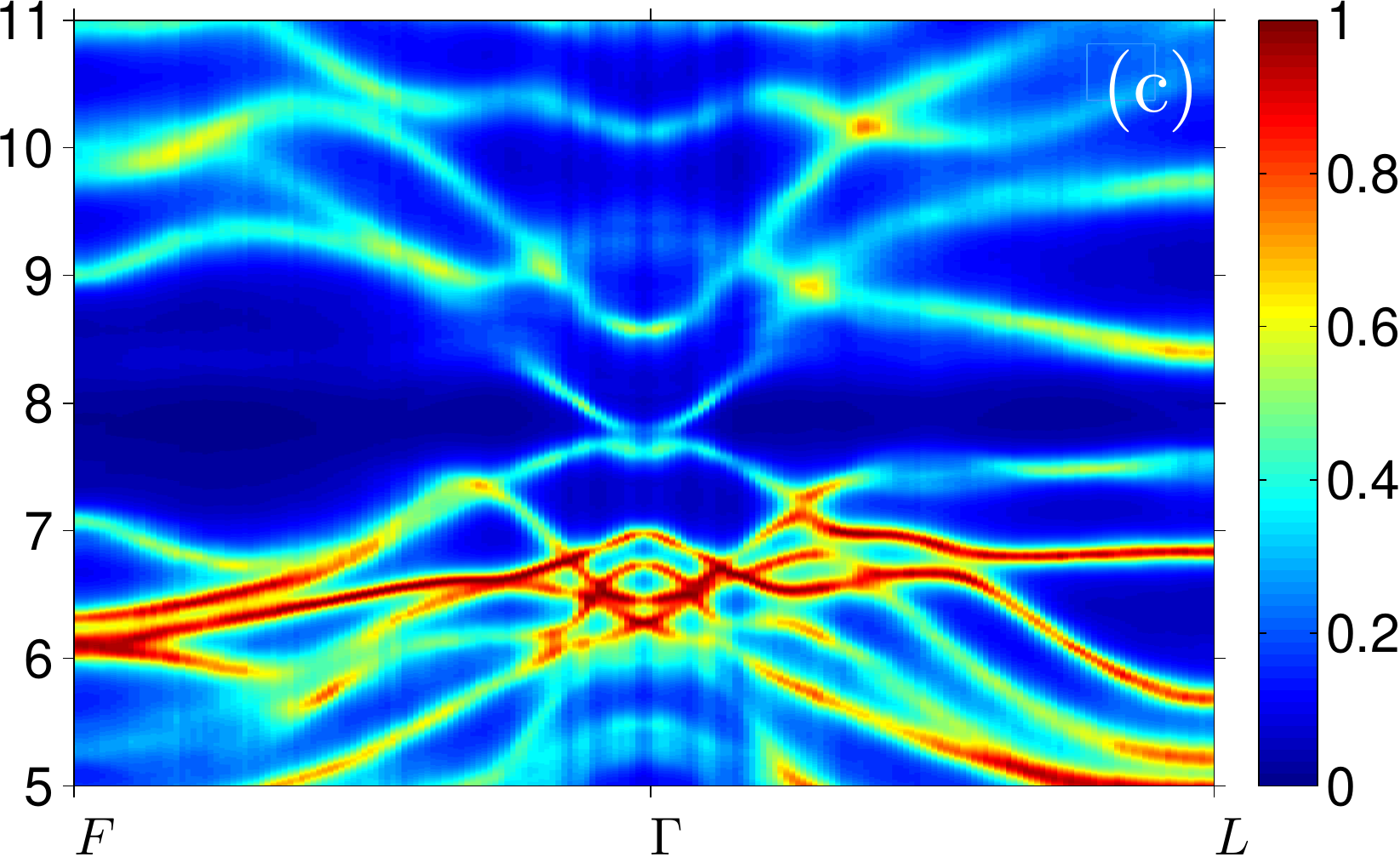}}
\end{minipage}%
\vspace{-0.1cm}
\begin{minipage}[c]{0.5\textwidth}
\subfigure{\includegraphics[height=3.5cm,clip]{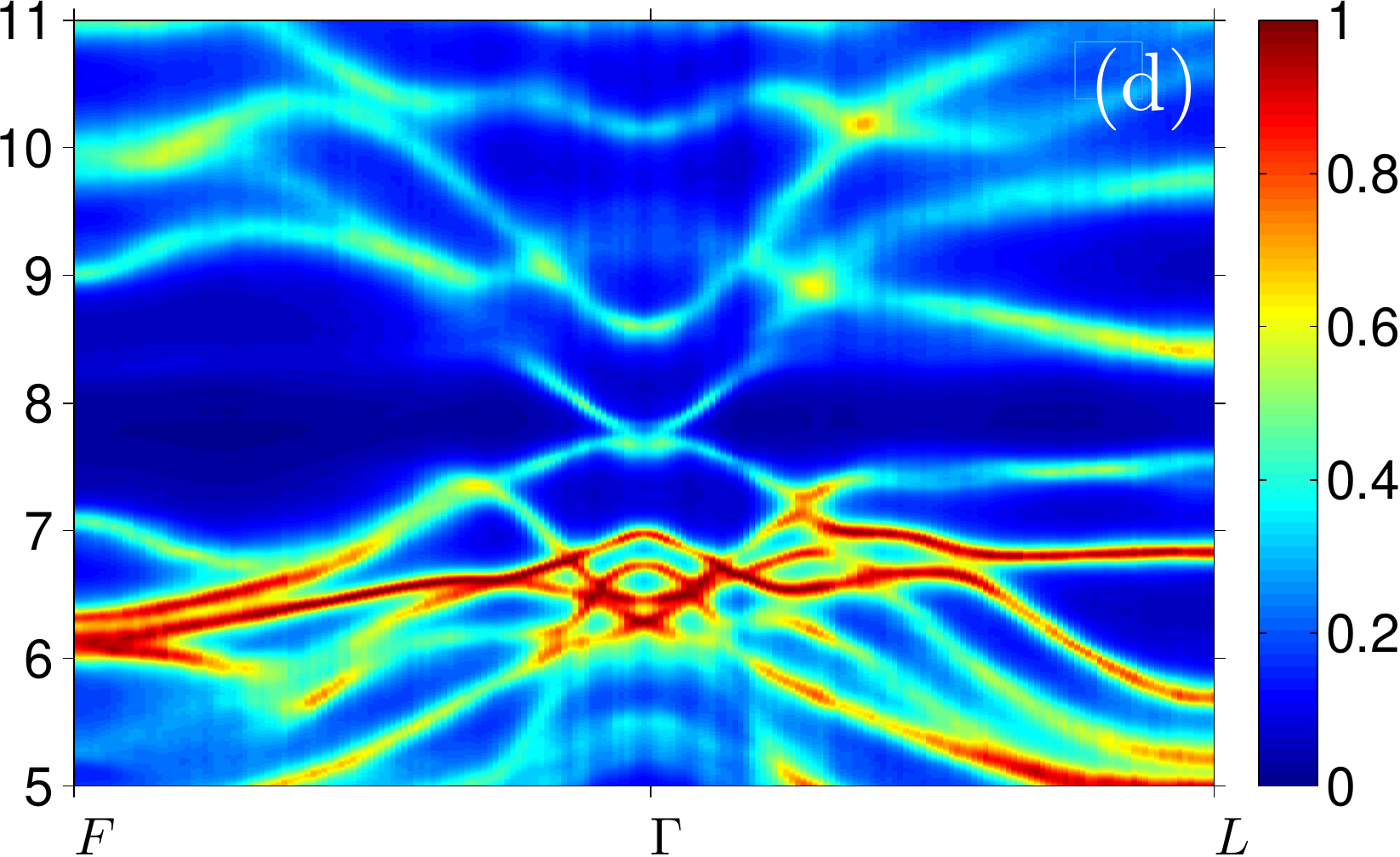}}
\end{minipage}%
\vspace{-0.1cm}
\begin{minipage}[c]{0.5\textwidth}
\subfigure{\includegraphics[height=3.5cm,clip]{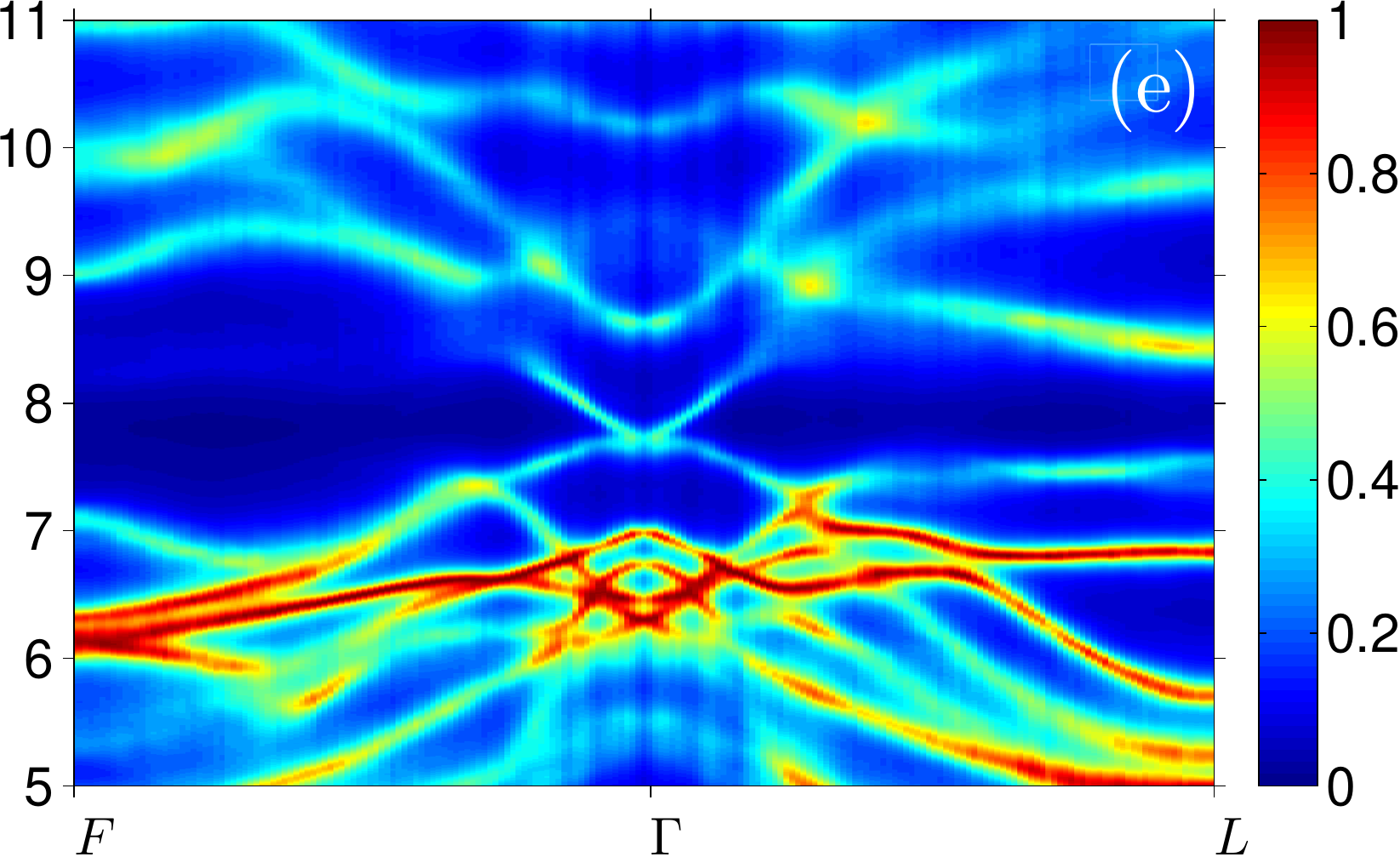}}
\end{minipage}%
\caption{Disordered spectral functions of \biin\ unfolded into
the primitive-cell BZ. (a) $x\!=\!8.3$\%. (b) $x\!=\!12.5$\%.
(c) $x\!=\!14.6$\%. (d) $x\!=\!16.7$\%.
(e) $x\!=\!18.8$\%.}
\label{fig:In-spectral} 
\end{figure}

The disordered spectral functions of \bisb\ are shown in
Fig.~\ref{fig:Sb-spectral},
with an artificial Lorentzian broadening of 2\,meV. At $x\!=\!68$\%, the
spectral gap is visible, but already very small, suggesting that
the system is approaching the critical point.  As $x$ is
increased to 78\%, a sharp Dirac cone is observed, which
remains robust from 78\% to 83\%. At
$x\!=\!88.5$\%, the band gap reopens, meaning that the system is in the NI
phase, and this topologically trivial phase becomes more robust as
$x$ goes to 99\% with a more visible gap.
One may notice that the effect of disorder is weak during
the phase-transition process,
and the semi-metallic behavior at criticality is rather sharp.

The spectral functions of \bisb\ at the $\Gamma$
point (in the primitive-cell BZ) are plotted in
Fig.~\ref{fig:Sb-spectral-Gamma}. At $x\!=\!68$\% there are
two peaks around the Fermi level, indicating that the CBM
and VBM are still separated, and the critical point has not been
reached yet. At $x\!=\!78$\% and 83\%,  the two peaks from the
conduction and valence bands merge into one, suggesting the system
becomes a semimetal.  As $x$ goes to 88.5\%, the gap opens up
again. From these results it appears that there is
a kind of ``critical plateau'' for $x$ between $\sim$78\% and $\sim$83\%.

This critical behavior observed for \bisb\ is at variance with
the general expectation for the topological phase-transition
behavior in the \bise\ class of TIs, where the system becomes a 
critical Dirac semimetal at one point in the 
parameter space (here it is the impurity composition $x$)
and then becomes insulating again immediately after the gap closure.
This deserves discussion.

We cannot exclude the possibility that numeric uncertainties
play a role here.
In \bisb, the band gap varies quite slowly as a function of $x$,
with a change of impurity composition of $5\%$ corresponding to
a band gap change of only $\sim$0.03\,eV. Thus, the critical point
could be hidden by disorder and artificial smearing, such that the system
looks metallic even if a very small band gap has opened up.
Moreover, finite-size effects may be important.
In a disordered system we expect that the
localization length evaluated in the middle of the mobility gap
should grow as the mobility gap shrinks and the system approaches
the critical metallic state.  As one approaches the critical point
at which this mobility gap vanishes,
the localization length may exceed the size of the supercell and
the states in neighboring supercells may overlap 
and behave like extended states.

However, it is also possible that a finite window of
metallic phase is physically correct.
Since the topologically nontrivial and trivial insulating
configurations compete with similar weight near criticality,  
the system may remain in the metallic phase until one of the two
insulating phases comes to dominate. Support for this
picture can be drawn from Ref.~\onlinecite{leung12}, in which
careful numerical simulations on a disordered lattice model showed
a finite-width region of metallic phase as the system was driven
from the TI to the NI phase with increasing disorder strength
while other parameters were held fixed.
In our case the disorder strength remains approximately
constant, but the ratio of disorder strength to energy gap varies
with $x$, so that a metallic plateau may still be expected.
We leave these questions
as avenues to pursue in future research.

The disorder-averaged spectral functions for \biin\ with the same
artificial broadening are plotted in Fig.~\ref{fig:In-spectral}.
At 8.3\%, the band-inversion character is still obvious
(note the tilde-like shape of the highest occupied bands around
$\Gamma$) with the spectral gap unclosed, which implies the system
may still stay in a TI phase. At
12.5\% the spectral gap becomes hard to recognize. When it comes
to 14.6\%, 16.7\% and 18.8\%, the spectral gap is almost completely
unrecognizable, which means the system is pretty close to the critical point.
Moreover, in sharp contrast with the behavior of \bisb,
the effect of disorder in \biin\ is very strong.
It can be seen from Fig.~\ref{fig:In-spectral} that
the original energy bands are strongly smeared out.
Different Bloch states are mixed together, as would be expected if
localized eigenstates are formed centered on the substituted
In atoms. It is difficult to identify the critical point simply by looking
at the disordered spectral functions, because the Dirac
semi-metallic behavior is not as obvious as in \bisb. Therefore,
we calculate the \Z2\ index of each configuration from $x\!=\!8$\% to
18.8\%. By inspecting the statistical behavior of the resulting
\Z2\ indices, as described next, we conclude that the critical point of
\biin\ is around 17\%.

Our theoretical prediction of $x_c$ for \biin\ is somewhat
higher than the experimental values, estimated to be 3\%-7\%
according to Brahlek \textit{et al.}\cite{oh-prl12} and $\sim$6\%
according to Wu \textit{et al.}\cite{armitage1} We attribute
our overestimate of $x_c$ to both the use of standard DFT
methods and the absence of impurity-impurity correlation terms
in Eq.~(\ref{equa:Heff}). Regarding the former, we would expect to get
an $x_c$ more consistent with the experimental results if hybrid
functionals or more advanced many-body first-principles methods
were used in the calculations, which unfortunately becomes
expensive for large supercells. Regarding the latter, we expect that the In
clustering effects would be treated more accurately if we would go
beyond a simple superposition of one-body impurity potentials and
include many-body terms in the impurity cluster expansion when
constructing the effective Hamiltonian.  However, this too would
carry a large computational cost due to the anisotropic nature of
the two-body impurity-impurity interactions and the fact that
higher-body terms may also be important.

\subsection{\Z2-index\ statistics}
\label{sec:Z2 stat}

The \Z2\ indices of a 3D band insulator are well defined for
a perfect periodic lattice with time-reversal symmetry. For disordered
systems, however, the topological indices are much harder to
calculate. A promising approach is the use of non-commutative
algebra,\cite{bellissard94,prodan11,teufel13,leung13}
but to date this has generally been applied to simple models,
and its applicability to realistic disordered materials has
not been demonstrated.

Here we attempt to determine the topological indices of a
disordered time-reversal invariant insulating system using a more
straightforward approach: we calculate the strong \Z2\ index
(with periodic boundary conditions on the supercell) for
each impurity configuration in the statistical ensemble, thus
determining the topological properties of the disordered system
from a statistical point of view. As long as the configurations
are sampled in such a way that each contributes
equally to the statistical ensemble, then we define the system
as \Z2-odd (i.e., a strong TI) if over half of the configurations
are \Z2-odd, and normal otherwise.
As mentioned in Subsec.~\ref{sec:spectral}, the impurity
configurations of \biin\ are generated using the Metropolis
Monte Carlo method based on a Boltzmann weight defined by the
In-clustering energy. As a result, the tendency of In segregation
is reflected in the number of generated distributed vs.\ clustered
configurations, rather than by manual assignment of weights.
Thus we consider each configuration to be equally weighted,
satisfying the criterion stated above.

The strong \Z2\ index statistics of several \biin\ configurations are
shown in Table~\ref{table:Z2-stat}.
For $x$ between 16.7 and 18.8\%, the number of \Z2-odd\ configurations
drops from eleven to five, so we estimate $x_c$ to be approximately 17\%.

\begin{table}
\caption{\Z2\ statistics\ of \biin. The entries in the first,
second, and third rows indicate respectively the number of
\Z2-odd, \Z2-even, and metallic configurations drawn from a
16-member ensemble.}
\centering
\begin{tabular}{l c c c c c c }
\hline\hline
           & 8.3$\%$ & 12.5$\%$ & 14.6$\%$ & 16.7$\%$ & 18.8$\%$  \\ [0.5ex]
\hline
\Z2\ odd &  15 & 16 & 14 & 11 & 5  \\
\Z2\ even & 0  & 0 & 0 & 3 &  9 \\
Metallic &  1  & 0  & 2 & 2 &  2 \\ [1ex]
\hline
\end{tabular}
\label{table:Z2-stat}
\end{table}
%

\section{summary and outlook}

To summarize, we have studied the topological phase transitions in
\biin\ and \bisb\ using two approaches, the direct application of
first-principles calculations on small supercells, and a Wannier-based
modeling approach that allows for a more realistic treatment of
disorder in large supercells.  Based on the former approach, the
$x_c$ of (Bi$_{1-x}$In$_{x})_2$Se$_3$ is slightly less than $12.5\%$,
while that of \bisb\ is even above $87.5\%$. A VCA treatment of
\bisb\ predicts $x_c\sim 65\%$; this is not in perfect agreement with
the prediction from the supercell calculations, but both of them are
much higher than that of \biin.  From the results of realistic
disordered calculations, we found that $x_c$ is $\sim$17\% for
\biin, while it is $\sim$78-83\% for \bisb.  The critical
concentrations are determined from disorder-averaged spectral
functions and \Z2-index\ statistics. It is concluded that in \bisb,
the band gap at $\Gamma$ decreases almost linearly with increasing
$x$, corresponding to the reduction in average SOC strength, with
only a very weak disorder effect.  For \biin, on the other hand,
the In $5s$ orbitals tend strongly to suppress the topological band
inversion even at very low impurity concentrations, so that the
phase transition is drastically accelerated as a function of
increasing $x$.

In the case of \bisb, we observed a critical plateau
from  $x\sim 78\%$ to $x\sim 83\%$. As discussed in
Subsec.~\ref{sec:spectral}, it is difficult to
say whether this intermediate metallic phase is just an
artifact of numerical limitations such as finite-size
effects, or is a true feature of the physics.  Further theoretical
and experimental work is needed to clarify what happens
in this critical region.

We also find a tendency of the In (but not Sb) atoms to segregate.
This In clustering effect could help clarify some aspects of the
topological phase transition in \biin, as for example by suggesting
a scenario in which the phase transition may happen locally, instead
of homogeneously as in the usual linear gap-closure picture. One
can imagine that as In atoms are implanted into bulk \bise, isolated
In clusters would start to emerge, inside which the system is
topologically trivial.  As more and more Bi atoms are substituted
by In, these isolated In ``islands'' become connected to each other,
and the topological phase transition happens when the percolation
threshold is reached.

Our results for \biin\ provide a physical explanation for the 
observed low transition concentration in several recent experiments
on \biin,\cite{armitage1,oh-prl12} and the results on \bisb\
may give predictions for future experimental works.

The techniques used in this paper provide a powerful methodology
that may be used to carry out theoretical explorations of other
types of disordered topological systems.  For example, interesting
physics is anticipated in a TI whose bandstructure is mostly
contributed by $p$ orbitals while substituting with impurities
having $d$ or $f$ orbitals. It could also be interesting
to investigate the effects of magnetic impurities, not only
on the surface states and their spin textures,\cite{henk1,henk2}
but on the bulk topological transition as well.
We thus hope that these methods will enable the search for new
materials and systems with non-trivial topological properties in
strongly disordered alloy systems.

\acknowledgments

This work is supported by NSF Grant DMR-10-05838. We are
grateful to M. Taherinejad and K. Garrity for useful discussions,
I. Souza for important technical assistance with the codes, and 
W. Wu and his collaborators for sharing their unpublished
experimental data.

\bibliography{doped_bi2se3}

\begin{thebibliography}{10}
\bibitem{comment_unfolding}%
  \BibitemOpen
  \bibinfo {note} {Expressed under Wannier basis, the summation over
  $\mathbf{K}$ in Eq.~(\ref{equa:unfolding}) becomes unnecessary, because there
  is a Kronecker-$\delta$ relationship between $\mathbf{k}$ and $\mathbf{K}$ in
  the expression for $\langle\psi_N(\mathbf{K})|\psi_n(\mathbf{k})\rangle$,
  thus removing the summation.}%
  \bibAnnoteFile{Stop}{comment_unfolding}%
\end{thebibliography}

\end{document}